\documentclass{article}

\usepackage{graphicx}
\usepackage{float}
\usepackage[utf8]{inputenc} 
\usepackage[T1]{fontenc}    
\usepackage{hyperref}       
\usepackage{url}            
\usepackage{booktabs}       
\usepackage{amsfonts}       
\usepackage{nicefrac}       
\usepackage{microtype}      
\usepackage{enumerate}      
\usepackage{subcaption}
\usepackage{amsmath}
\usepackage{color}

\newcommand{\beq}{\begin{equation}}
\newcommand{\eeq}{\end{equation}}

\title{A Model for Image Segmentation in Retina}
\author{Christopher Warner$^{1,2}$ and Friedrich T. Sommer$^{1,3,4}$\\
$^1$Redwood Center for Theoretical Neuroscience\\
$^2$Biophysics Graduate Group \\
University of California, Berkeley \\
Berkeley, CA 94720\\
$^3$Helen Wills Neuroscience Institute \\
University of California, Berkeley\\
Berkeley, CA 94720\\
$^4$Intel Labs, Santa Clara, CA 95054-1549\\
\texttt{cwarner@berkeley.edu, fsommer@berkeley.edu}
}

\begin{document}
	\maketitle

\begin{abstract}
 While traditional filter models can reproduce the rate responses of retinal ganglion neurons to simple stimuli, they cannot explain why synchrony between spikes is much higher than expected by Poisson firing \cite{brivanlou1998mechanisms}, and can be sometimes rhythmic \cite{neuenschwander1996, koepsell2009}. Here we investigate the hypothesis that synchrony in periodic retinal spike trains could convey contextual information of the visual input, which is extracted by computations in the retinal network.
 We propose a computational model for image segmentation consisting of a Kuramoto model of coupled oscillators whose phases model the timing of individual retinal spikes. The phase couplings between oscillators are shaped by the stimulus structure, causing cells to synchronize if the local contrast in their receptive fields is similar. In essence, relaxation in the oscillator network solves a graph clustering problem with the graph representing feature similarity between different points in the image. We tested different model versions on the Berkeley Image Segmentation Data Set (BSDS). 
 Networks with phase interactions set by standard representations of the feature graph (adjacency matrix, Graph Laplacian or modularity) failed to exhibit segmentation performance significantly over the baseline, a model of independent sensors. In contrast, a network with phase interactions that takes into account not only feature similarities but also geometric distances between receptive fields exhibited segmentation performance significantly above baseline. 
\end{abstract}

\section{Introduction}

For decades the commonly accepted view of retinal processing has been that it provides a bank of independent, linear filters that decorrelate stimulus features in space and time, reducing the redundancy in the retina's representation \cite{barlow1961}. Linear spatio-temporal filters factorized into center-surround spatial and biphasic temporal components followed by pointwise non-linearities encode local stimulus features in the spike rates of retinal ganglion cells (RGC) \cite{kuffler1953}. There remain, however, severe puzzles, unexplained by the textbook view of retina.

First, for retinal ganglion cells it would be inefficient to use spikes exclusively in a rate code with rather long integration window.   This assumption is in conflict not only with theoretical principles, such as the efficient coding hypothesis \cite{atick1992}, but with experimental observations.  
For example, it has been shown that time to first spike in RGCs can be very reliable, containing nearly as much information about the stimulus as spike rates \cite{gollisch2008}.

Second, the circuitry in the anatomical retinal network is exquisitely complex, consisting of >60 distinct neuron types stratified into at least 12 parallel and interconnected circuits providing roughly 20 diverse representations of the visual world, discussed at length in \cite{masland2011}, \cite{masland2012a}, \cite{werblin2011}, \cite{gollisch2010}. Simple linear spatio-temporal filtering requires only a handful of cell types in the outer retina, leaving the rest of the network unexplained. By "Occam's razor", the textbook view must be at least incomplete.

Third, the textbook model of retina fails to account for complex phenomena such as precise spiking of RGCs relative to the phase of network oscillations in the gamma range (50-80Hz) \cite{neuenschwander1996, koepsell2009}. Although the function of retinal oscillations is yet unknown in mammals, they have been observed in mouse \cite{menzler2011}, cat \cite{neuenschwander1999} and primate \cite{ogden1973}. Further, gamma-band retinal oscillations have been causally connected to the perception of spatially extended stimuli in the frog \cite{ishikane2005}.  Specifically, it has been observed that neurons in the cat lateral geniculate nucleus (LGN) often receive periodic retinal spike trains in the gamma band. Estimates of information rate in LGN spike trains suggest that in cells with periodic inputs, the spike train could mulitplex two different types of information. While rate modulation in a courser time window encodes local stimulus contrast, a significant fraction of the total information is encoded by spike timing at a fine time scale, conveying the phase of the gamma frequency in the neurons input \cite{koepsell2009}.


Fourth, computational models reflecting the text book view, such as the linear nonlinear Poisson (LNP) model and generalized linear models (GLM), predict RGC responses to a simple white noise stimulus \cite{schwartz2006} with reasonable accuracy. However, looking more closely, one observes pairwise correlations in retinal activity, even in the absence of stimulus (correlations) \cite{schneidman2006}. Taking into account these pairwise activity correlations improves decoding of retinal responses to white noise \cite{pillow2008} -- but does not explain why the retina introduces such correlations to begin with. The situation with ecologically relevant natural movie stimuli, in which pixels possess dependencies across space and time, is even more puzzling. The model prediction by independent encoding models becomes rather poor \cite{schwartz2006}, and even encoding models that include second-order correlations fail to replicate responses to natural movie stimuli \cite{chichilnisky2016}. We suggest to take these mismatches between retina and its current computational models as an encouragement to design and investigate novel computational models of retina.

Here we approach the challenge to design better retina models from a computational perspective and ask:  "What type of image analysis could be computed in an array simple sensors with access to (center surround) image features, like found in retina, above and beyond independent sensors proposed in the textbook model?"  Specifically, we follow the lead suggested in the discussion of experimental work \cite{ishikane2005} and investigate whether, in addition to encoding local image features, the retinal network can also extract spatially extended visual features and multiplex the extracted information into the retinal output using phase synchrony in periodic spike trains \cite{koepsell2009}.

To concretely design a sensor network model with this function, we build on contributions provided in various streams of earlier work, the insight that image segmentation (IS) can be cast as a graph clustering problem \cite{shi2000}, and the insight that, in addition to spectral methods, graph clustering can be efficiently solved in networks of phase-coupled oscillators \cite{arenas2006}. To evaluate the performance of the model, the Berkeley Image Segmentation Dataset (BSDS) was essential. While the motivation for this work is to model a computation in retina, it should be noted that the network model we propose is still quite abstract. The model aims to serve as a proof of principle that the network computation could be efficiently performed by biological retinas, it is not intended as a neurobiolocically detailed circuit model.

\begin{figure}[!h]
    \centering
    \centerline{\includegraphics[scale=0.33]{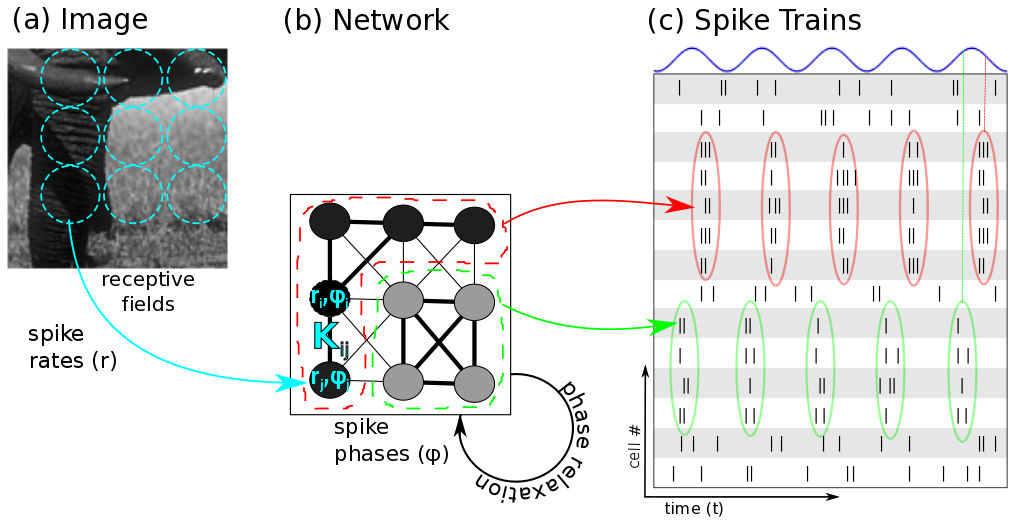}}
    \caption{ \small{ \textbf{Image Segmentation Model:} (a) Input image with superimposed retinal receptive fields (dashed cyan circles). (b) Network of retinal neurons. The neural firing rates $r_i$ represent local contrast in the receptive fields. The phase interactions $K_{ij}$ are displayed by the links between neurons. Line thickness represents the strength of the interaction which is set by the similarity of local features. Recurrent propagation in the network produces the phase structure $\phi_i$ of the periodic spike trains. (c) Resulting spike trains. Information about local features is represented in firing rates and segmentation is represented in phase structure.} }
    \label{SpikeTrains}
\end{figure}

A coarse overview of the model is given in Fig.~\ref{SpikeTrains}. The firing rate $r_i$ in a coarse time window represents the local image contrast in the classical receptive field of neuron $i$. The similarity between pairs of local features in the image determines the strength of phase interaction between the periodic structure in the spike trains. Phase diffusion through the phase couplings does not change firing rates but produces sets of neurons with similar spike times on a fine time scale. These sets of synchronous neurons represent spatially extended image features, image segments. The resulting spike trains multiplex two types of information, local contrast in individual spike rates, and image segments in sets of neurons that fire nearly synchronously \cite{koepsell2009}. In our example, two image segments are represented by groups of neurons with different phases. Note that in this study, we only consider models of the phase dynamics, omitting aspects of spikes and spike rates.

The remainder of this paper is structured as follows. The Methods section describes prerequisites for our study from the literature. Section~\ref{BSDS} concisely defines the putative computation of our retina model, image segmentation (IS) using simple image features available in retina, local contrast values or local center surround image features. The evaluation pipeline proposed in the BSDS image segmentation database \cite{martin2001} is explained, which is essential to quantitatively compare the performances of different models. Following \cite{shi2000}, section~\ref{GraphClustering} describes how image segmentation can by cast as a graph clustering problem, and how an adjacency graph is constructed for a particular image. Section~\ref{GC_methods} describes three common graph clustering methods from the literature, average association, graph Laplacian and modularity, that we will compare in our image segmentation experiments. Section~\ref{Kuramoto} describes how, as an alternative to computing eigenvalues of the graph representation matrix, relaxation of phase-coupled oscillators can be used to solve graph clustering problems. This step is critical in mapping the computation of image segmentation to the network model in Fig.~\ref{SpikeTrains}b.

The Results section contains original contributions of our study. Section~\ref{results_TopoModu} describes topographic modularity, a novel graph-clustering method based on modularity \cite{newman2006} that we propose for clustering multigraphs. Image segmentation can be understood as clustering of of a multigraph, in which one type of edges represent feature similarity and the other geometric vicinity of the features in the image plane. Section~\ref{Results_KurNet} compares the performance of image segmentation of commonly used eigenvector-based "spectral methods" \cite{chung1997} for graph clustering to the method of phase relaxation \cite{arenas2006}. We find that phase relaxation generally outperforms spectral methods, independent of the choice of a particular image graph or receptive field structure. Thus, our further experiments focus on phase relaxation, the method that also has the advantage of being easily implementable as an oscillation-based computation \cite{koepsell2010}. The central experimental results of our study are described in section~\ref{GaussRF}. We compare segmentation performances of different network models to a baseline segmentation algorithm based on thresholding image feature histograms, a computation which does not require a network. While the standard graph clustering methods are not able to significantly outperform histogram thresholding, one model stands out significantly, the network implementing topographic modularity. Section~\ref{Recall} describes experiments to elucidate why the network with topographic modularity outperforms the competitor models. 


In the Discussion section we delineate the various implications of the presented results. We describe the predictions our model makes for future neuroscience experiments and its potential for applications of image processing with coupled sensors.


\section{Methods}

    \subsection{Berkeley Segmentation Data Set} \label{BSDS}
	
\begin{figure}[!h]   
          \centering
          \centerline{\includegraphics[scale=0.33]{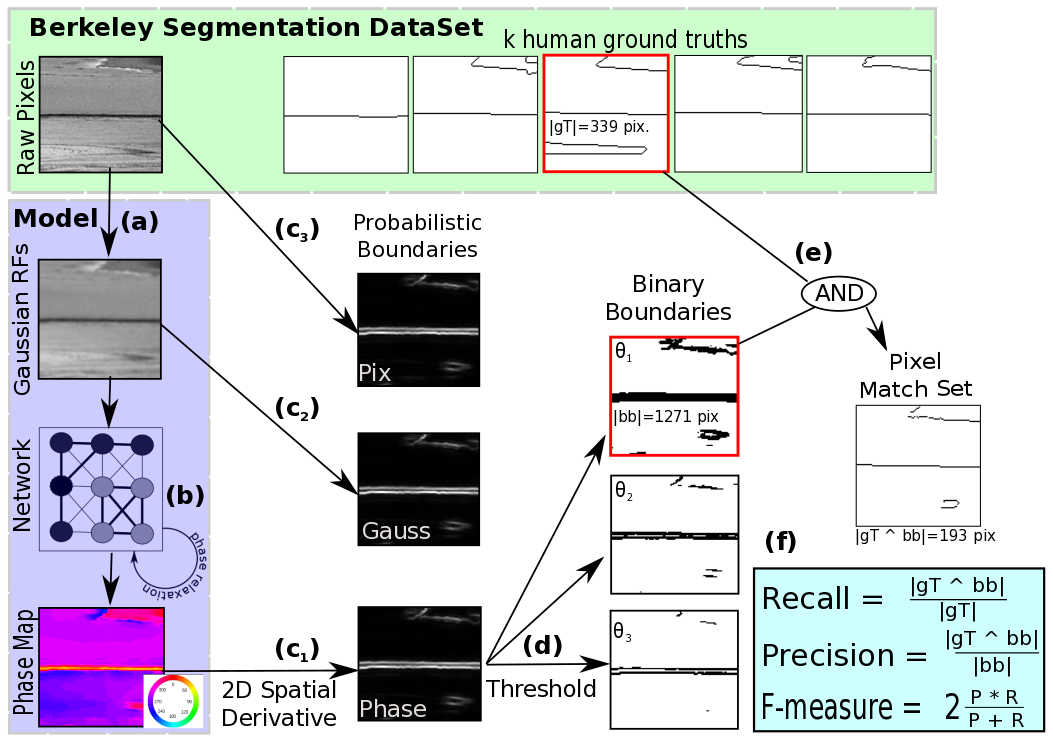}}
          \caption{ \small{ \textbf{ Performance and benchmarking:} Input image patch and associated human drawn ground truth boundaries (gT) provided by BSDS is displayed in the green box. The operations performed by model are displayed in the blue box. Other steps of model evaluation are illustrated in the remainder.  
          (a) Filtering the raw image patch with a Gaussian kernel ($\sigma=1$). 
          (b) Phase relaxation in the network (Fig.~\ref{SpikeTrains}b) produces a phase map. 
          (c~$_{1}$) Spatial gradient operation ($\nicefrac{\delta}{\delta r}$) and normalization resulting in probabilistic boundary map ($pb \in [0,1]$).
          (d) Thresholding pb map at several values yielded binary boundary (bb) maps. 
          (e) Match set was computed for each bb-gT pair at different distance tolerances, $d_t$.
          (f) Precision, recall and F-measure were computed by ratios of boundary pixel sets. 
          (c~$_{2,3}$) To assess the performance of network models relative to baselines, we repeated steps (c) - (f) on Gaussian RF and image pixels independent sensors models, comparing F-measures by subtraction.} }
        \label{Pipeline} 
    \end{figure}

    Image segmentation is a challenging and important problem in computer vision and the Berkeley Segmentation Data Set (BSDS) is a standard benchmarking data set for many computer vision image segmentation algorithms \cite{martin2001, arbelaez2011}. It consists of 500 large ($\sim 400 \times 300$ pixels) color images  each with multiple ($\sim 5$) human drawn boundary contours (green box in Fig.~\ref{Pipeline}), as well as code provided for standard benchmarking and comparison of algorithms. Since image segmentation is closely related to boundary detection and quantification of boundary detection performance is more straightforward than that of image segmentation, segments in images are often recast as boundaries for benchmarking. Binary boundary pixel locations are compared to human drawn boundaries using the precision, recall, f-measure framework. In this context, "Precision" is the proportion of image pixels hypothesized by a method to belong to segment boundaries that agree with the ground truth. "Recall" is the percentage of ground truth boundary pixels that are found by a method. F-measure is the harmonic mean of Precision and Recall.

    In order to leverage the BSDS resource, we must first convert the output of a segmentation model - a phase, spectral or feature activation map (blue box in Fig.~\ref{Pipeline}) - into binary boundaries. Intuitively, a good segmentation of an image has been achieved if the model output map has very similar values within segments and large discontinuities at boundaries. We compute spatial derivatives ($\nicefrac{\delta}{\delta r}$) in the output map and normalize the values between 0 and 1, allowing us to interpret resulting probabilistic boundary (pb) as the algorithm's confidence that there is a boundary between segments at a particular image location. We can threshold pb's at multiple values and compare each resulting binary boundary map (bb) to each human drawn ground-truth boundary map (gT), generating a pixel match set by a logical AND operation. Because human drawn boundaries are not precise down to the pixel, we allow small misalignment between gT and bb pixel including a pair in the match set if they are within $d_t$ pixels of one another.

     We compared the ability of different phase coupled oscillator models to segment images from the Berkeley Segmentation Dataset (BSDS) \cite{martin2001}. The models differed in the phase couplings. One baseline model contained isotropic couplings, while the couplings in the other models AA, GL, M and TM were the transformations of the adjacency of features described above. We set the parameters $\sigma_f$, $\sigma_d$ and $\sigma_\omega$ to adequate common values and performed for each method a parameter grid search in neighborhood connectivity $R_{M}$ and scale $K_{s}$ to maximize the average $\Delta F_b$ across 500 image patches. The oscillator frequency was 60 Hz (typical for retinal gamma oscillations) and we gave the networks 300ms to relax the phases, corresponding to an interval between saccades.

    \subsection{Image segmentation as a problem of graph clustering} \label{GraphClustering}
    
    Within a stage of visual processing, in which a set of local visual features is extracted, image segmentation can be viewed as a graph clustering problem \cite{shi2000}. Consider an image and its corresponding neural representation in retina or LGN, in which the activity in individual cells represent the strength of local center-surround features. Image segmentation consists of clustering sets of local image features that share properties and thus likely correspond to larger objects in the image. However, much more efficient than clustering pixel values, is to apply clustering on more sophisticated local image features, e.g., center-surround, edges, multi-scale textures, as done in state-of-the-art segmentation methods \cite{shi2000,sarkar1998}.   

    The problem of graph clustering, that is finding “cliques” of strongly connected nodes in a graph, is obviously related to finding pixel sets with similar local features. To recast image segmentation in terms of graph clustering, one first uses kernels to construct an adjacency matrix, in which an element is large whenever two features have similar values and lie nearby each other in the image plane \cite{shi2000,sarkar1998}. The segmentation of the image corresponds to finding the communities (subsets of nodes) that are strongly interconnected within the community, and well separated from nodes outside. The goal then is to find non-trivial subsets of nodes that can be separated from one another by cutting through the minimum weight of edges, know as the "mincut" problem. Though a brute force, optimal solution to this problem would be combinatorically intractable, approximate solutions can be found efficiently by leveraging the machinery of "spectral graph theory" \cite{chung1997}.
    
    Following \cite{shi2000}, we define a graph for segmenting an image by the \textit{ adjacency} matrix:
    \begin{equation}
        \mathbf{A}_{ij}  = e^{-\frac{ { (f_i - f_j)^2  }  }{ 2 \sigma_{f}^2 } } \cdot e^{ -\frac{(r_i - r_j)^2  }{ 2 \sigma_{r}^2 } } \cdot  \bigg(1-H(\sqrt{(r_i - r_j)^2}-R_M)\bigg) 
    \label{adjacency}
    \end{equation}
    \noindent with $H(x)$ the Heaviside step function. The first factor reflects the dissimilarity of the local features $f_i$ and $f_j$, in our case local contrasts. It was found experimentally that $\sigma_{f}=0.2$ provides reasonable dynamic range in adjacency weight distribution. The second and third terms reflect the distance between the local features in the image plane. Since we are interested how well segmentation can be performed in networks with local neighborhood connectivity and for simplicity, we we null out the second term by setting $\sigma_r = \infty$ and add the third term, a binary rectangular Heaviside function 1 - $H(\sqrt{(r_i - r_j)^2}-R_M)$ that is 1 within a maximum radius, $R_M$, and 0 outside. We explored $R_M$ values of 1,3,5 and 10.

    \subsection{Three common graph clustering methods} \label{GC_methods}

    The simplest strategy of graph clustering, referred to as \textit{average association} (AA), is to analyze the adjacency matrix directly \cite{sarkar1998}.  Eigenvalues of the adjacency quantify the amount of correlated structure and the associated eigenvectors characterize the location of the correlated structure in the image.  Other methods of graph clustering utilize transformations of the adjacency matrix, often incorporating the node "degree", $d_i = \sum_j A_{ij}$, which captures the total weight of connections to each node from all other nodes in the network. One such transformation we considered is the normalized \textit{ graph Laplacian} (GL) or Kirchhoff matrix:  $L = D^{-1/2} (D - A) D^{-1/2}$ with diagonal matrix $D_{ij}= \delta_{ij}\sum_k{A_{kj}}$, $\delta_{ij}$ the Kronecker symbol.  This strategy, combined with more sophisticated image features, forms the basis of a very successful image segmentation algorithm, the ``Normalized Cut'' \cite{shi2000}. The eigenvectors and associated smallest eigenvalues of the Laplacian matrix find divisions in the input characterized by large feature differences.
    
    A third transformation of the association matrix we considered is \textit{modularity} (M) \cite{newman2006}, which has successfully discovered community structure in social and information networks, outperforming the graph Laplacian in these tasks. The modularity matrix can be written as
    \begin{equation}
        Q = A - N
        \quad \text{with} \quad 
    	N_{ij} = D_i D_j
        \quad \text{and} \quad 
    	D_i := \frac{d_i}{\sqrt{2m}}
    	\label{eq_mod}
    \end{equation}
    
    \noindent where $D_i$ and $D_j$ denote the ``degree'' of nodes $i$ and $j$ respectively, normalized by the total weight of edges in the graph, $2m = \sum_k{d_k}$.  Importantly, the null model matrix, $N$, contains the expectation of the weight value between each node pair $N_{ij}$ based on the strength of connectivity of both nodes. In this way, an expected graph is constructed by assuming an otherwise random graph with node degrees constrained (an Erdos-Renyi random graph). Comparing the observed adjacency graph to the null model by subtraction reveals graph structure beyond what could expectedly be introduced by heterogeneous node degrees. In section~\ref{TopoModu}, we discuss modularity further and introduce an extension, called \textit{ topographic modularity} (TM), with null model adapted for graphs embedded in space.
    
     Once an associated matrix representing a graph is constructed, spectral methods have been predominantly used within the graph clustering community to find clusters within because eigenvalues and eigenvectors efficiently find an approximate solution to the combinatorially intractable "mincut" problem. It has been observed on simple networks that the eigenvalue spectrum of an associated matrix resembles the temporal progression of clusters discernible from phases of nodes in a Kuramoto network \cite{arenas2007}, this time evolution of clusters forming a hierarchical clustering of a network. Given this observation, we compute the time evolution of a phase coupled oscillator network dynamical system as an alternative to eigenvector-based graph clustering methods.

    \subsection{Kuramoto Phase Relaxation Model} \label{Kuramoto}
    
    The described graph clustering methods in \ref{GraphClustering} compute the eigenvectors of the associated matrices \cite{chung1997} which, in essence, is assessing anisotropic diffusion in these networks. This process has also been related to the path a random walker would take through the graph where edge weights represent transition probabilities and the distribution of electrical potentials on nodes in a resistor network where an edge weight represents the conductance of a particular resistor \cite{grady2006}. A further parallel has been between eigenvector based methods for graph clustering and the "fundamental mode(s) of a spring-mass system" \cite{shi2000}. To rigorously investigate this last claim, we simulate phase relaxation in a network of Kuramoto coupled oscillators \cite{kuramoto1984} with networks defined by methods described in \ref{GraphClustering}.

    Here we followed \cite{arenas2007} and assessed diffusion properties by relaxing a network of phase-coupled oscillators : 
    
    \begin{equation}
    \Delta \phi_i = \omega_i + \sum_j {K_{ij} sin(\phi_i - \phi_j)}
    \text{,} \quad \quad
    K_{ij} = k_s M_{ij}
    \label{kur}
    \end{equation}
    
    \noindent with each node's natural frequency $\omega_i=60Hz$ and where $M_{ij}$ is one of the graph matrices mentioned above. For intuition, Eq.~\ref{kur} loosely simulates a lattice of oscillating masses connected by different size and signed springs. The lattice is shaken at initialization and through the relaxation dynamics, masses connected by strong positive springs are attracted in phase while strong negative springs repel one another. In the original Kumamoto model \cite{kuramoto1984}, couplings $K$ were set to be uniform, supporting isotropic diffusion. As a baseline, we also investigated the effects of isotropic diffusion (ISO) for image segmentation. Unlike the uniform network, a network with heterogeneous weights relax to stable states containing multiple distinct clusters of phase aligned oscillators. 
    
    In the implementation of the model, the overall positive scaling factor $k_{s}$ is critical. If coupling weights are too large, phasers will spin wildly in response to even small phase differences. Conversely, if too small, oscillators will adjust their phase too slowly and the relaxation will not converge in time. Importantly, the phase relaxation was limited to 300ms or 20 periods of the 60Hz signal, which is the average duration of fixation before a saccade brings the eye's gaze to a new point, refreshing the input and beginning the computation once again. The value for the $k_s$ parameter was set for each graph individually based on mathematical considerations in equation \ref{kur}. A middle value $k_s^{ \tiny{mid} }$ was chosen so that the phase change of the node with largest degree $D_{k_{max}}$ is limited to $\nicefrac{\pi}{2}$ radians in one full period of the 60Hz signal when all its neighbors are aligned $\nicefrac{\pi}{2}$ radians away and exerting maximal pull.
    
    \begin{equation}
    k_s^{ \tiny{mid} } = 60Hz \cdot \frac{2\pi}{\nicefrac{\pi}{2}} \cdot D_{k_{max}}
    \end{equation}
    
    \noindent We then bracketed that value above and below by an order of magnitude. 
    
    The final result of the phase relaxation simulation is a phase map with a phase value, $\phi_i \in [0,2\pi]$, associated with every node, $i$, in the network and corresponding location $i$ in the image. Spectral methods also yield a value associate with each location, $i$, in the image with $v_i \in [-\infty, \infty]$. In order to compare our results to other algorithms using the BSDS resources, we convert these maps to probabilistic boundaries and recast the image segmentation problem as a boundary detection one as discussed in Section \ref{BSDS} and illustrated in Fig.~\ref{Pipeline}.

\begin{figure}[H]
        \begin{subfigure}[t]{0.48\textwidth}
            \centering \includegraphics[scale=0.2]{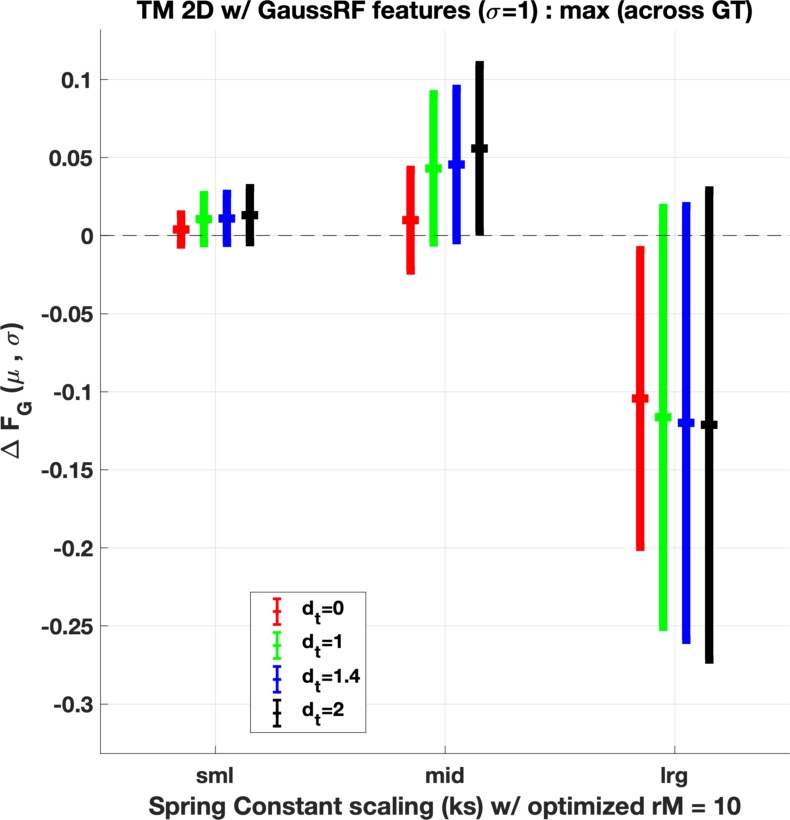}
            \caption{}
        \end{subfigure}
        ~
        \begin{subfigure}[t]{0.48\textwidth}
            \centering \includegraphics[scale=0.2]{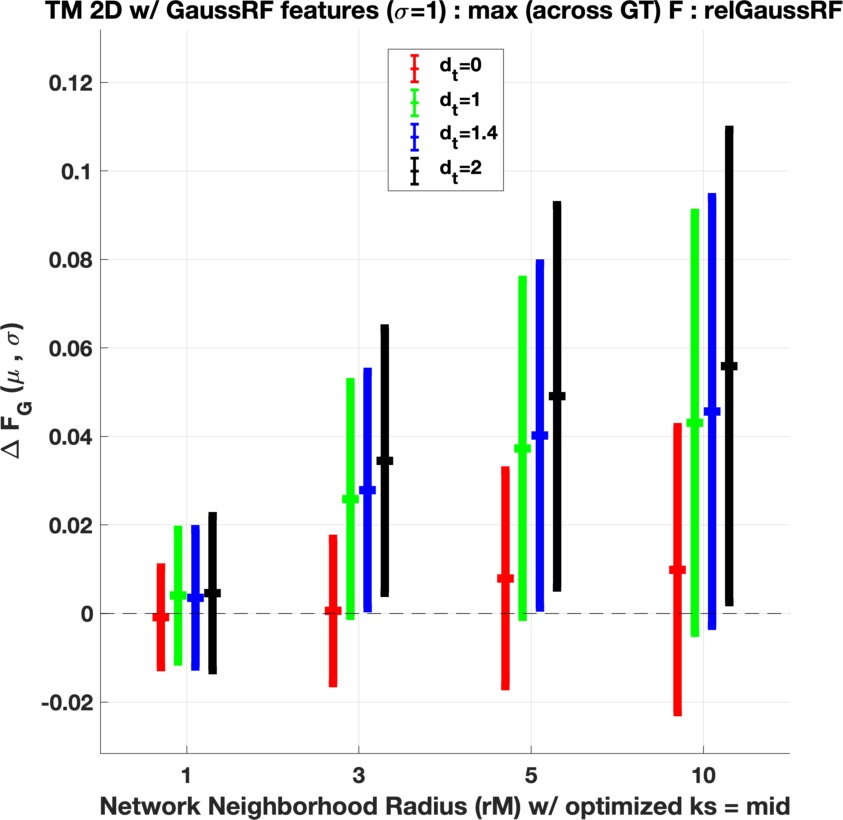}
            \caption{}
        \end{subfigure}
        \caption{ \small{ \textbf{ Hyper-parameter optimization:} Network neighborhood graph structure $r_M$ and coupling spring-constant scaling $k_s$ are important meta parameters of the algorithm, discussed in Sections \ref{GC_methods} and \ref{Kuramoto} respectively. We plot mean and standard deviation across 500 image patches of $\Delta$F-measure relative to Gaussian RF independent sensors for the 2D topographic modularity network. Colors indicate pixel distance tolerances $d_t$ (see Fig.~\ref{Pipeline} for explanation). \emph{Left} panel shows performance at three $k_s$ values, with $r_M$ fixed at optimal. \emph{Right} panel shows performance at four $r_M$ values, with $k_s$ fixed at optimal. Fig.~\ref{rM_ks_oneImg} shows the effects of the different parameters on a single example image patch. } }
        \label{rM_ks_stats}
    \end{figure}

\begin{figure}[H]
        \centering
        \includegraphics[scale=0.25]{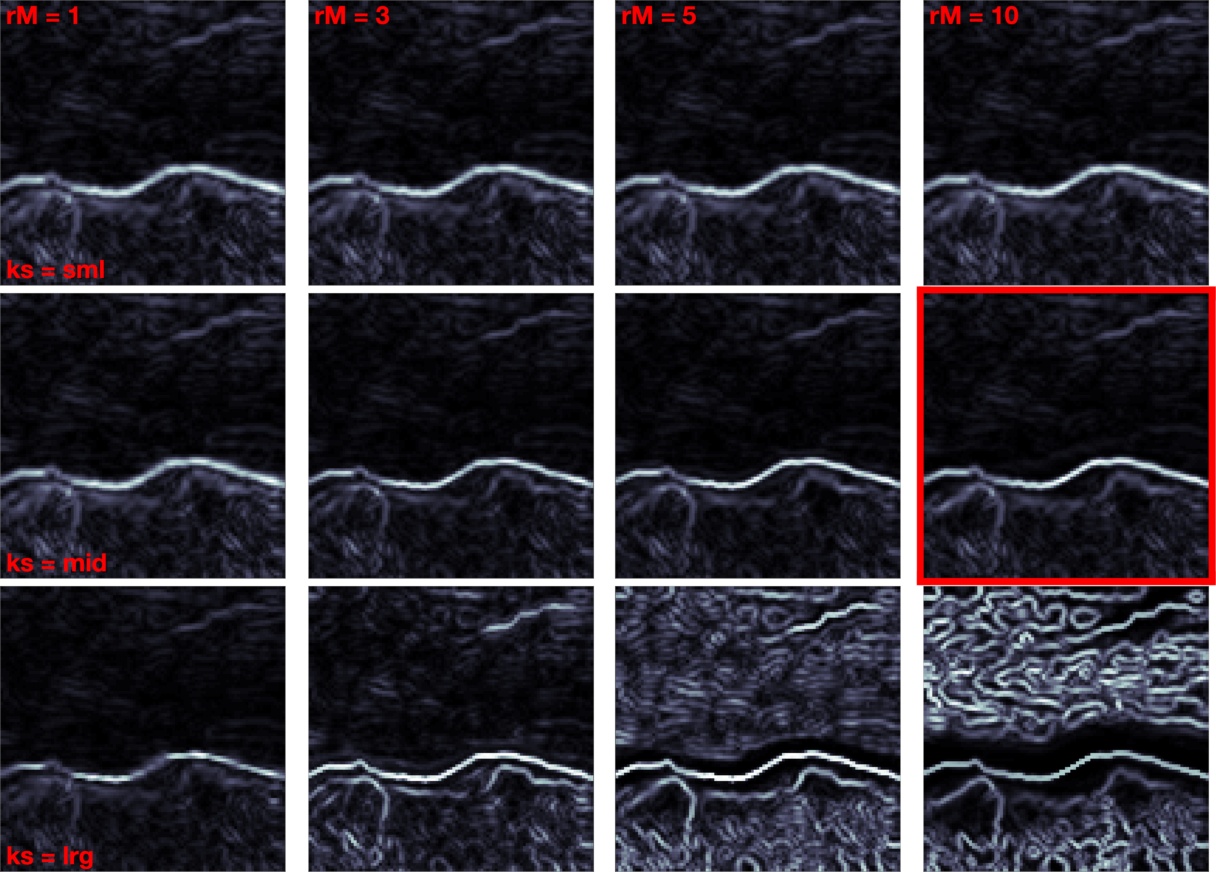}
        \caption{ \small{ \textbf{ Effect of hyper-parameters, single image patch example: } Probabilistic boundary maps shown for resulting phase distribution from TM 2D method for combinations of 3 $k_s$ (rows) and 4 $r_M$ (columns) hyper-parameters. }  }
        \label{rM_ks_oneImg}
    \end{figure}

    In practice, two meta parameters, $r_M$ defining the neighborhood structure of the Adjacency graph and $k_s$ defining an overall scaling on the strength of phase interactions in the network, impacted image segmentation performance. They were optimized for each method and results shown are with optimized parameters, shown in Fig.~\ref{DelF_STD}. To optimize parameters for each method, we performed segmentation of 500 image patches with four $r_M$ values ranging from 1 to 10 and bracketing $k_s$ as discussed above and chose the parameter settings with best average performance across all images and across $d_t$. Fig.~\ref{rM_ks_stats} illustrates the procedure for one particular method. It shows average performance across $k_s$ values for optimal $r_M$ on the left and performance across $r_M$ values for optimal $k_s$ on the right. Fig.~\ref{rM_ks_oneImg} shows the effect of different parameter settings on one example image patch.
    

\section{Results}

    \subsection{Modularity null models for images} \label{results_TopoModu}

    An image can be described by a multi-graph, in which pixels or local image features are represented by nodes and each pair of pixels has two different types of edges connecting them. One edge type represents geometric distance in the image plane and the other edge type represents feature differences. The two types of edges are given by adjacency matrices, resulting from the two types of distances and corresponding kernel functions (like a Gaussian kernel), as in Eq.~\ref{adjacency}. Shi and Malik \cite{shi2000} proposed a way to collapse this multi-graph of an image to an ordinary graph by forming the Hadamard product of the two adjacency matrices. An entry in the resulting single adjacency matrix $A$ represents the two distinct similarities between pixels, geometric proximity and feature similarity by a single number. Specifically, an entry in $A$ can only be large, if both, distance and feature differences are small in the corresponding pair of pixels. In order to find image segments, researchers then used "spectral" graph clustering methods on the matrix $A$ \cite{sarkar1998, shi2000}.
    
    For some graph clustering methods, such as modularity \cite{newman2006}, the collapsing of the multi-graph into an ordinary graph destroys information, which is critical for segmenting images. The modularity matrix consists of the difference of the adjacency matrix and a null model. The null model represents an average adjacency value.  
    In the standard modularity method, Eq.~\ref{eq_mod}, the average is computed from the degrees of the two nodes involved, the row and column sum of the collapsed graph. 
    However, in natural images, the average feature similarity of a pair of pixels is a function of geometric distance \cite{ruderman1994}, see also Fig.~\ref{Adj_vs_dist} in supplemental section \ref{temporal_null_model}.  Thus, an appropriate null model for images should also depend on the geometric adjacency matrix.
    
    \begin{figure}[H]   
        \centering      
        \centerline{\includegraphics[scale=0.47]{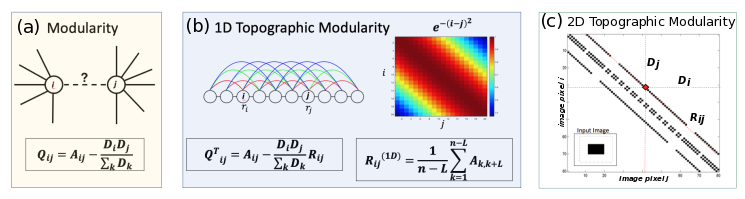}}
        \caption{ \small{ \textbf{ Modularity null models \& space}:  In the null model of Newman's modularity \cite{newman2006} \textit{ (panel a)}  the average weight between nodes $i$ and $j$ is proportional to the product of their node degrees ($D_i \cdot D_j$). The topographic modularity's null model \textit{ (panels b \& c)} additionally includes a distance-dependent factor, $R_{ij}$, which is the average edge weight between all node pairs in the graph separated by the same distance that separates nodes $i$ and $j$. \textit{ Panel b} illustrates $R_{ij}$ for a schematized 1D graph, shown with edges colored based on distance between the nodes they connect. Inset plot shows geometric factor in the topographic null model. 
        Each term in $R_{ij}^{(1D)}$ is an off-diagonal sum in the adjacency matrix. \textit{ Panel c} shows the mask associated with a single geometric distance in a 2D image. Here $R_{ij}^{(2D)}$ at 1 pixel separation has a complex structure in the Adjacency matrix for even the simple binary image shown in the inset.} }
        \label{TopoModu}
    \end{figure}

    To address this issue, we devised a novel graph clustering method called \textit{ topographic modularity} (TM) in which the null model takes topographic 
    distance in the image plane into account. Like the standard modularity \cite{newman2006}, see Fig.~\ref{TopoModu}a and Eq.~\ref{eq_mod}, an entry of the topographic modularity matrix, $Q^T$, is the difference between the entry $A_{ij}$ and the expected connectivity, captured by the null model, $N_{ij}$. Here, the topographic null model $N_T$ accounts for distance dependent factors in feature similarity with the $R_{ij}$ term in addition to node degree heterogeneity.
    
    \begin{equation}
    	Q^T_{ij}  = A_{ij} - N^T_{ij} 
        \quad \text{where} \quad 
    	N^T_{ij} = D_i \cdot D_j \cdot R_{ij}
    \end{equation}
    The $R_{ij}$ factor represents the average connectivity between all node pairs that are separated by the same geometric distance as the nodes $i$ and $j$. For a network in space along a 1D line, Fig.~\ref{TopoModu}b, the distance dependent contribution to the null model can be written mathematically as 
    \begin{equation}
    	R_{ij}^{(1D)}  = (\frac{1}{n-L})\sum_{k=1}^{n-L}{A_{k,k+L}}
    	\label{NM-1D}
    \end{equation}
    where $L$ is the distance separating nodes $i$ and $j$ (i.e., $L = r_i-r_j$) and $n$ is the total number of nodes (or pixels or features in the image). In 1D, the average connectivity of all nodes separated by a distance L is equal to the mean along the L'th diagonal.
    
    For networks with 2D grid-like geometry, like Adjacency graphs constructed from images, the computation of $R_{ij}^{(2D)}$ is more involved, yet the interpretation is the same. Reshaping a 2D image into a 1D vector so that similarity relationships can be represented in a 2D matrix introduces discontinuities in spatial relationships between entries in the matrix. Weights between nodes separated by a particular distance can be labeled by a mask specific to the dimensions of a particular image. 
    Fig.~\ref{TopoModu}c shows the weights between all neighboring nodes ($L=1$) in the network derived from the 11 x 11 binary image in the inset. For completeness, we show $R_{ij}^{(2D)}$ masks for other pixel separations in supplement section \ref{MotivateModularity} Fig.~\ref{2D_Bij}.

\begin{figure}[H]
      \centering
      \textbf{Null Model Consistency}\par\medskip
      \centerline{\includegraphics[scale=0.2]{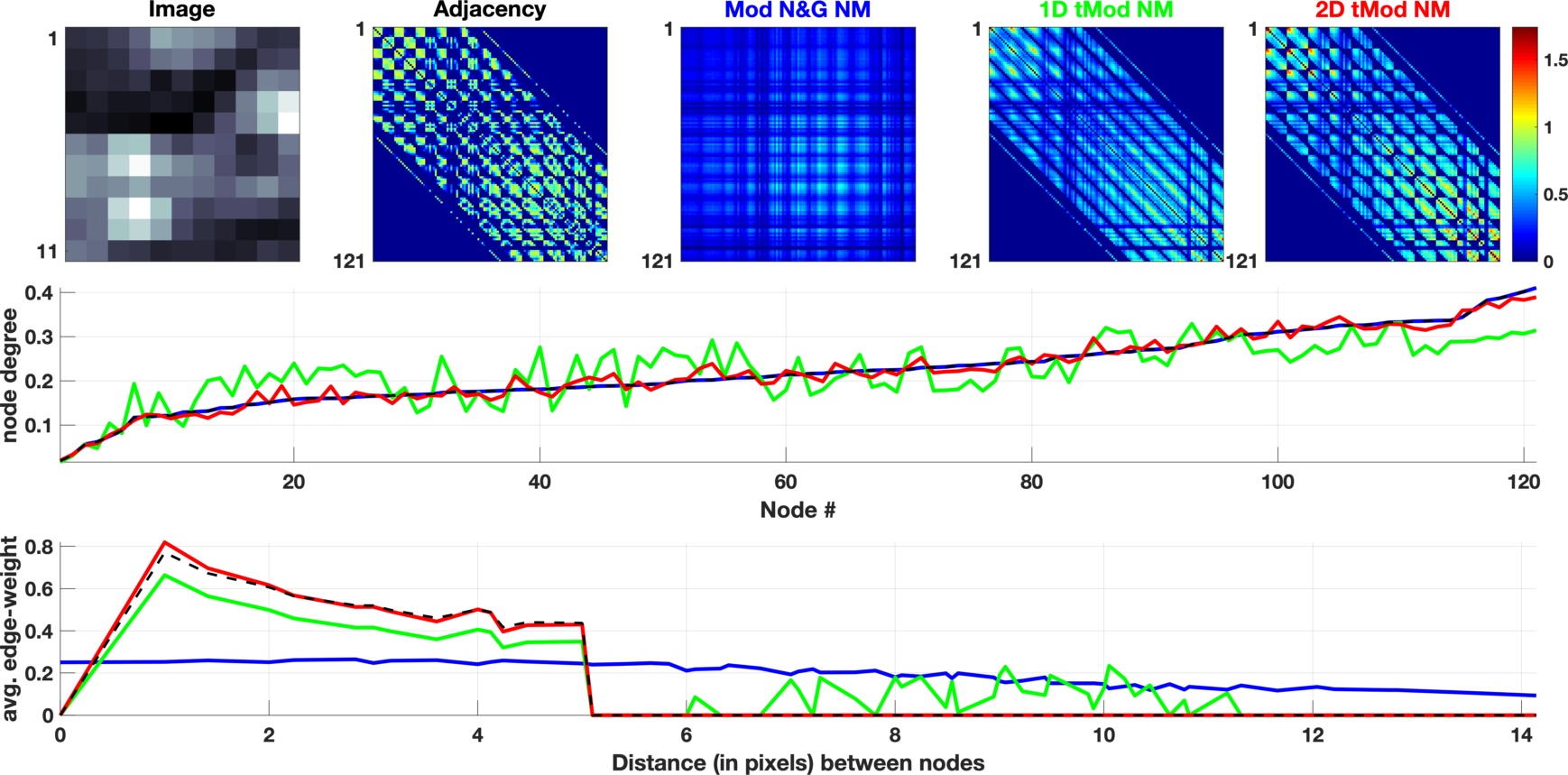} }
      \caption{ \small{ \textbf{ Null model consistency}: \textit{ Top row} from left to right shows image patch, the adjacency (black) constructed from the patch with $r_{max}=5$, and null models for modularity (blue), 1D topographic modularity (green) and 2D topographic modularity (red), with colorbar indicating edge weight. Models represented by line colors in plots as well. \textit{ Center plot} shows average node degree (row sums in each matrix) sorted by strength in adjacency. \textit{ Bottom plot} shows average edge weight as a function of distance in the image plane.} }
       \label{Mod_NM_consistency}
    \end{figure}

    Before comparing the different null models in an image segmentation we compare how well they capture the structure of an adjacency matrix of an image. The null model in Newman's modularity, by construction, is a "consistent" estimator of node degrees \cite{chang2012}, ensuring that $\sum_j{N_{ij}} = \sum_j{A_{ij}}$ (blue line in Fig.~\ref{Mod_NM_consistency} middle). However, it is clearly the wrong null model for natural image Adjacency graphs for two reasons. First, the null model incorrectly contains positive diagonal weights in proportion to $D_i^2$, although the diagonal elements of the adjacency matrix are zero. Second, it does not capture the distance dependence of the adjacencies, thereby
    underestimating average adjacency between proximal nodes and overestimating it for distant node pairs. 
    Both problems manifest in the difference between the blue and the dashed lines in Fig.~\ref{Mod_NM_consistency} bottom. 
    
    While the TM-1D and TM-2D null models are not strictly consistent in node degree or distance dependence, they are nearly so (green and red lines respectively in Fig.~\ref{Mod_NM_consistency}). Introducing distance-dependent statistics into the TM-1D null model corrects for the spatial "inconsistency", vastly improving estimates of edge-weight over M. TM-2D offers improvement over TM-1D due to further refinement of its null model, see Eq.~\ref{NM-1D} and surrounding text. Further discussion of null model consistency and bias in the supplemental section \ref{ModularityConsistency}. \\
    
    %
    
\begin{figure}[H]
        \centering
        \includegraphics[scale=0.30]{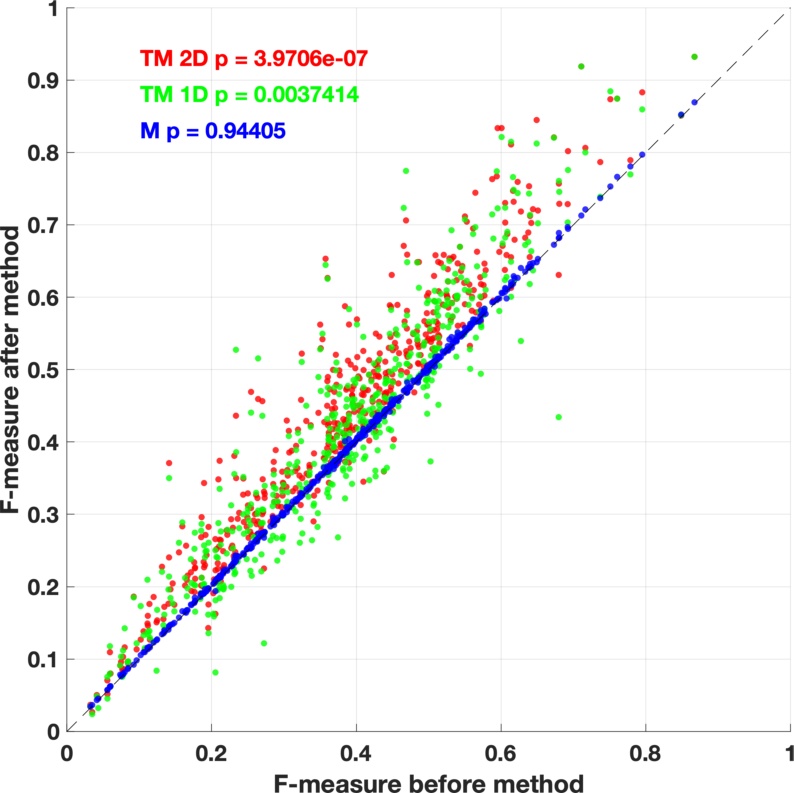}
        \caption{ \small{ \textbf{ Modularity performance comparison:} Each scatter point represents one image patch. Newman's modularity (M) in blue, 1-dimensional topographic modularity (TM 1D) in green and 2-dimensional topographic modularity (TM 2D) in red. Points above the unity line indicate image patches with improved image segmentation with network phase relaxation over-and-above Gaussian RF independent sensors. P-values quantify the difference between F-measure distribution across 500 image patches before and after network computation. } }
        \label{scatter_3mods}
    \end{figure}

    Importantly, the difference between an adjacency value and its average in the modularity can become negative. In a Kuramoto net relaxation, these negative weights mediate phase repulsion and introduce targeted phase desynchronization, see Sec.~\ref{Kuramoto}, at boundaries in an image where gross image statistics change. 
    In contrast, if the modularity value between a node pair is positive, it contributes to phase synchronization. 
    Fig.~\ref{scatter_3mods} illustrates image segmentation performance before and after phase relaxation through connections defined by M (in blue), TM-1D (in green) and TM-2D (in red). While M does not significantly change image segmentation performance over Gaussian RF independent sensors, TM-1D does so (p-value $\sim0.004$) and TM-2D does so even more (p-value $\sim 4 \cdot 10^{-7}$). With TM-2D, we see improvement for $\nicefrac{\sim 460}{500}$ image patches. \\
    

    \subsection{Broad comparison of models on image segmentation} \label{Results_KurNet}
    
    Following \cite{koepsell2010}, we investigate the idea whether a phase-coupled network of simple sensors of local image features, similar to those in the retina, could at the same time represent local and contextual image features in its output.
    Specifically, phase interactions mediated through heterogeneous network edges which are influenced by local features similarities can segment an image, grouping regions within a segment into the same relative phase and introducing phase breaks at segment boundaries. In a biological system, the contextual image information encoded by phase can be represented by the timing of spikes and be multiplexed into spike trains, whose rates represent the local features Fig.~\ref{SpikeTrains}.
    
    This idea is tested on images provided in the Berkeley Segmentation Data set (Sec.~\ref{BSDS}). For an image patch, we construct a graph based on local features in the image (Sec.~\ref{GraphClustering}) and segment the image by either computing eigenvectors or by performing anisotropic phase diffusion in a Kuramoto net. Computing a spatial derivative on either eigenvectors or the final phase distributions and normalizing values between 0 and 1 converts the output into probabilistic boundaries, which can be quantitatively compared to assess relative performance of different image segmentation methods.
    
    We ask whether a phase-coupled sensor array can add to an image segmentation that can be done based on the independent sensor measurements alone. Thus, the network computation must outperform two baseline methods. The first method computes normalized spatial gradients on the raw image pixels (magenta, RawPix). In the second method the image pixels are first convolved with Gaussian receptive fields, roughly similar to those measured in retina (cyan, GaussRF). As a third baseline method, we include isotropic diffusion in a network with homogeneous phase couplings between nearest-neighbor nodes (black, ISO).  
    
        \begin{figure}[H]
          \centering
          \centerline{\includegraphics[scale=0.4]{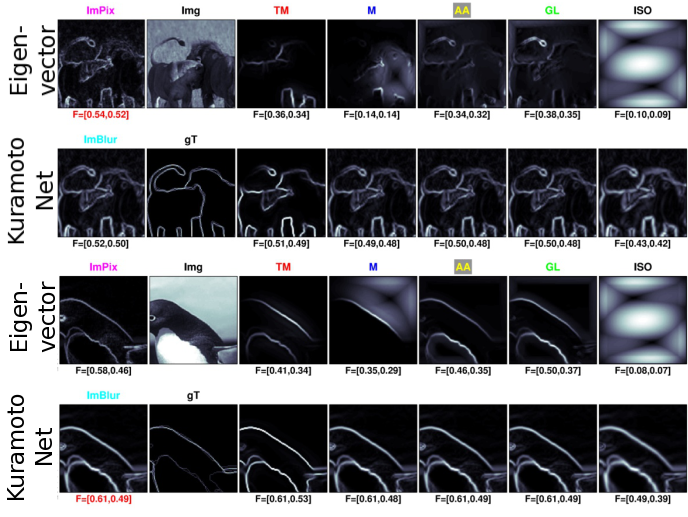}}
          \caption{ \small{ \textbf{ Spectral methods vs. Kuramoto Net Examples: } Two example image patches (top two rows and bottom two rows) show probabilistic boundaries found by different network (TM, M, AA, GL) and baseline models (ImPix, GaussRF, ISO). Network models are segmented using eigen-methods (1st and 3rd row) and Kuramoto Net phase relaxation (2nd and 4th row). Qualitatively, boundaries found with spectral methods are less crisp and more localized than those found with Kuramoto Net phase relaxation. } }
        \label{EigPbs}
    \end{figure} 

    Probabilistic boundaries (pb) can be interpreted as the algorithm's confidence that a boundary exists between two segments at a particular location in the image. Fig.~\ref{EigPbs} shows pb's resulting from the segmentation of different networks constructed from the same image patch, either by computing eigenvectors and by performing Kuramoto net relaxation. Qualitatively, we observe that eigenvectors seem to focus a spotlight on a region of the image patch while information propagated through the Kuramoto Net covers all parts of the image patch. Regardless of the network method used, boundaries found with the Kuramoto net are crisper and extend further across the image patch than do those found by computing eigenvectors.
    
        \begin{figure}[H]
          \centering
          \centerline{\includegraphics[scale=0.5]{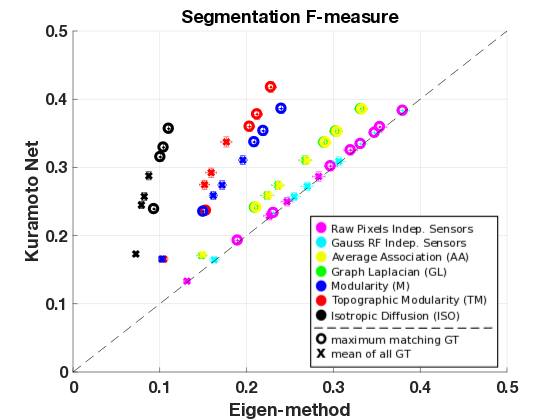}}
          \caption{ \small{ \textbf{ Spectral methods vs. Kuramoto Net Statistics:} F-measure computed across 500 image patches, mean and standard error errorbars. Colors indicating different network and baseline models are used consistently throughout this paper. Circles indicates that F-measure for each image patch taken for maximum matching GT and x's shows mean value across all GT's. Network models built with Gaussian RF features are segmented by the best combination of the top 3 eigenvectors on the x-axis and by the phase distribution after Kuramoto Net relaxation on the y-axis. The dashed unity line indicates equal performance and the independent sensors baseline models (magenta and cyan) do not deviate from it.} }
        \label{KurVsEig}
    \end{figure} 
    
    To assess whether this trend in image segmentation performance is statistically significant, we calculate Precision, Recall and F-measure across 500 image patches, shown in Fig.~\ref{KurVsEig}. Plotting F-measure statistics for network and baseline models segmented by Kuramoto-net and Eigen-methods, we find that segmentation without network computation (magenta and cyan) outperforms the results from the best combination of the top 3 eigenvectors, regardless of the model. We also find that all scatter points lie above the unity line, indicating superior image segmentation performance of anisotropic phase diffusion in a Kuramoto net verses the spectral clustering methods. As a consequence of this observation, we focus in the reminder on the superior methods based on Kuramoto Nets. \\

    \subsection{Influence of receptive fields choice} \label{GaussRF}
    
    We further observe that the features from which networks are constructed influence segmentation performance achieved. This comes as no surprise since state-of-the-art image segmentation algorithms rely on a combination of sophisticated spatially-extended features. 
    
\begin{figure}[H]
      \centering
      \centerline{\includegraphics[scale=0.47]{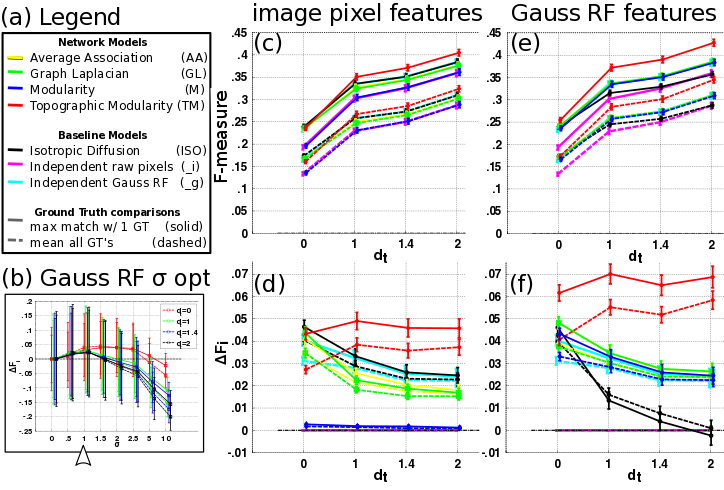}} 
      \caption{ \small{ \textbf{ Gaussian RFs improves segmentation}: Performances of 4 anisotropic diffusion and 3 baseline models are compared using raw image pixel features and Gaussian RF features, \textit{ center and right columns respectively}, lines representing average and bars standard error across 500 BSDS image patches. (\textit{ a}) Colors indicate different models and line styles indicate ground truth comparison as in Fig.~\ref{KurVsEig}. (\textit{ b}) Optimal spread, $\sigma$, of Gaussian RF's chosen by maximizing change in F-measure relative to the independent raw pixels baseline model, $\Delta F_i$, averaged across all image patches. Recall $d_t$ is the "distance tolerance" when computing the pixel match set, Fig.~\ref{Pipeline}. Optimal performance for all $d_t$ values obtained for Gaussian RF $\sigma=1$. (\textit{ c}) F-measure and (\textit{ d}) $\Delta F_i$ when models receive raw image pixels as features. (\textit{ e}) F-measure and (\textit{ f}) $\Delta F_i$ when models receive Gauss RF activation as input features.} }

    \label{BlurVsNo}
    \end{figure}

    We constrain our investigation to the relatively simple and local stimulus features that retina is supposed to have access to. Specifically, we investigate the difference in segmentation caused by switching between raw pixels and Gaussian receptive fields with different radii.   
    Again, we compare the segmentation performance of networks with phase relaxation to baseline models representing independent sensors, 
    and a model with isotropic diffusion through a homogeneous neigbor connections. 
    We find that Gaussian receptive field features provide better segmentation than raw image pixels both when used as independent sensors and to construct phase interaction networks. Fig.~\ref{BlurVsNo} shows the segmentation performance (F-measure and the change in F-measure relative to the independent sensors image pixels baseline model) as a function of pixel match distance tolerance ($d_t$).
    
    For small tolerances $d_t$ in the F-measure (see section 2.2) the simpler isotropic phase diffusion model was a surprisingly strong competitor, even beating some of the anisotropic networks (black lines in Fig.~\ref{BlurVsNo}c and d). Isotropic diffusion with optimized parameters provides mild smoothing of image structure, which operates indiscriminately within and across segments. To introduce the effect of smoothing in other models, we introduced Gaussian RF features. The filters corresponded to optical blur and the extended (centers of) receptive fields in retinal ganglion cells. Fig.~\ref{BlurVsNo}b shows segmentation performance as a function of the width of the Gaussian filter, $\sigma$, and tolerance parameter, $d_t$. We find that Gaussian RF features with $\sigma=1$ were beneficial and near optimal across different tolerance values. Interestingly, the size of the optimal Gaussian coincides with the size of retinal ganglion cell receptive fields measured in primate retina \cite{croner1995}. See supporting information \ref{Optimal_GaussRF} for further discussion.

Fig.~\ref{BlurVsNo}e and \ref{BlurVsNo}f show the improvement in segmentation performance using Gaussian features above using image pixel features. In particular the method TM displayed a significant increase in $\Delta F_i$ which became more prominent for larger pixel match distance tolerances $d_t$. Among all methods TM was able to improve segmentation performance the most, compared to that achievable with the Gaussian RF independent sensors model.

    \subsection{Detailed model comparison between most promising models} \label{Det_model_comp}

\begin{figure}[H]
      \centering
      \centerline{\includegraphics[scale=0.23]{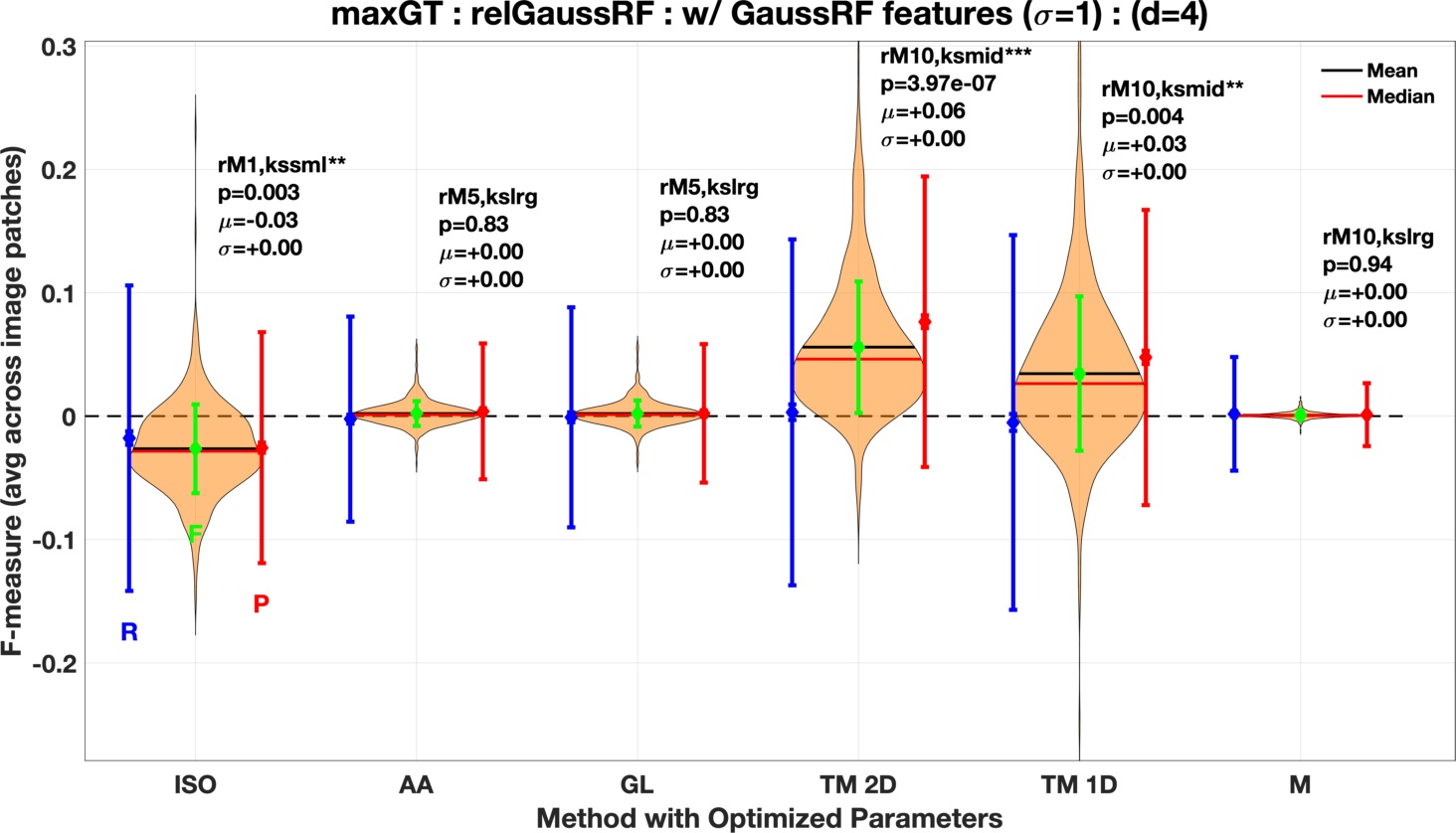}}
      \caption{ \small{ \textbf{$\Delta$F-measure model comparison:}  Violin plots show $\Delta F_G$ distribution with moments of $\Delta$Precision, $\Delta$Recall and $\Delta$F-measure distributions across 500 image patches in red, blue and green, respectively. $\Delta$F relative to Gaussian RF Independent Sensors model. Optimal hyper-parameters($r_M$,$k_s$), statistical significance, p-values and distribution moments indicated above each method. Performance of ISO, TM-1D and TM-2D models relative to Gauss RF are statistically significant, as determined by Mann-Whitney U (aka rank-sum) test.} }
    \label{DelF_STD}
    \end{figure}

    To assess the overall performance of different models on the diverse input images, each model was run with optimized parameters. Fig.~\ref{DelF_STD} shows image segmentation performance improvement from Gaussian RF independent sensors. Here the models TM-1D, TM-2D and ISO were significantly different from the three other methods that stayed near baseline $\Delta F = 0$. ISO stayed below baseline because the input kernels provide near optimal blur and therefore additional isotropic blurring deteriorated the segmentation performance. TM-1D performs well too, but not as well as TM-2D. This is because the null models are increasingly accurate, section \ref{results_TopoModu}. Shown results are with best matching ground truth. Results hold with average across all human drawn ground truths, though less pronounced. 

    
    Segmentation performance via anisotropic phase diffusion in a Kuramoto net depends critically on the structure of the phase couplings. Kuramoto nets using the graph Laplacian, average association or Newman's modularity as the phase couplings do not improve segmentation performance significantly over the independent sensors Gaussian RF baseline model. Only the Kuramoto model with the topographic modularity as phase couplings increases segmentation information over baseline independent sensors, homogeneous network and competitor heterogeneous network models, as quantified by the F-measure. \\
    
    \subsection{Why is the Kuramoto model with topographic modularity superior?} \label{Recall}

\begin{figure}[H]
        \centering
        \begin{subfigure}{.32\textwidth}
            \centering
            \includegraphics[width=\textwidth]{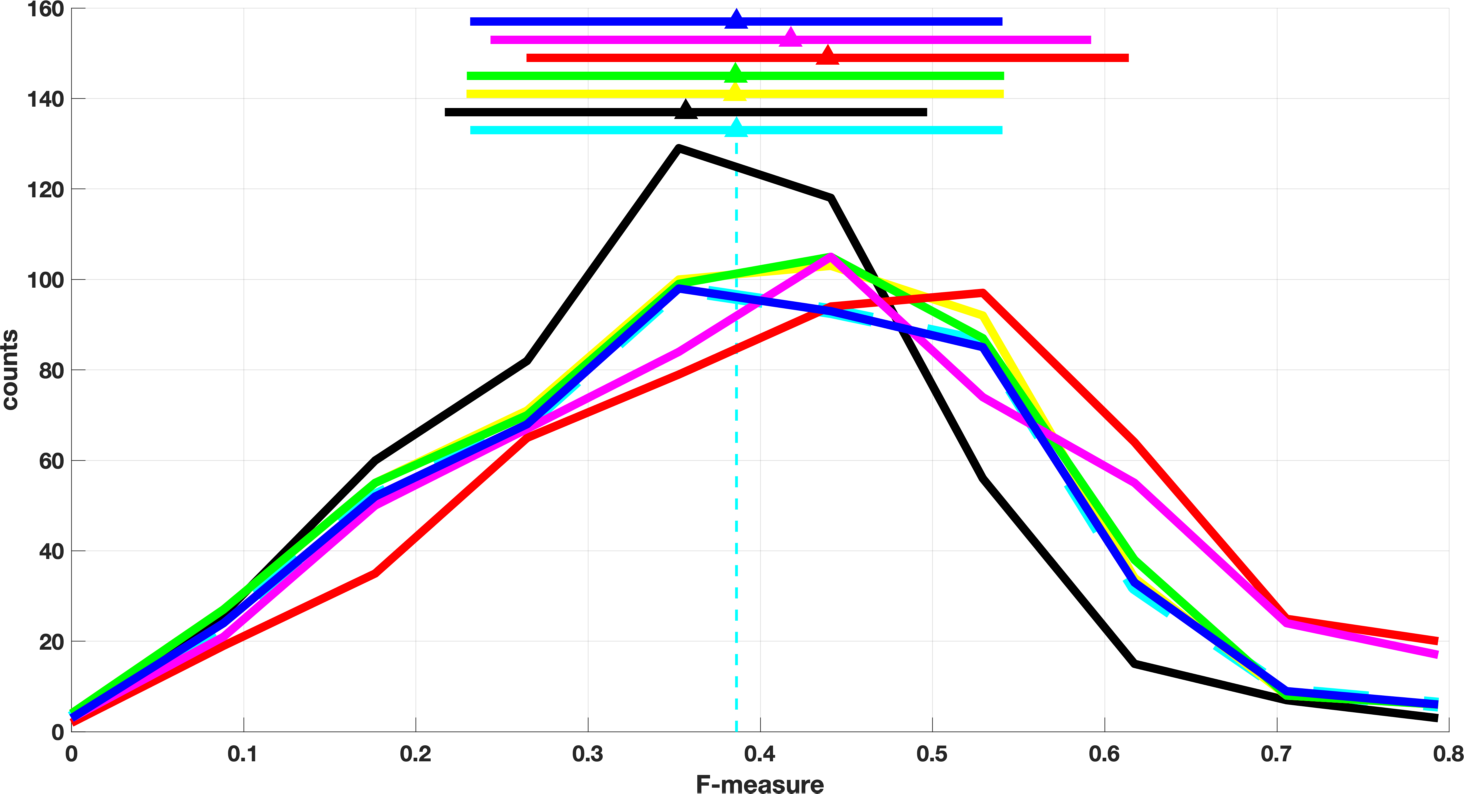}
            \caption{F-measure}
        \end{subfigure}
        \begin{subfigure}{.32\textwidth}
            \centering
            \includegraphics[width=\textwidth]{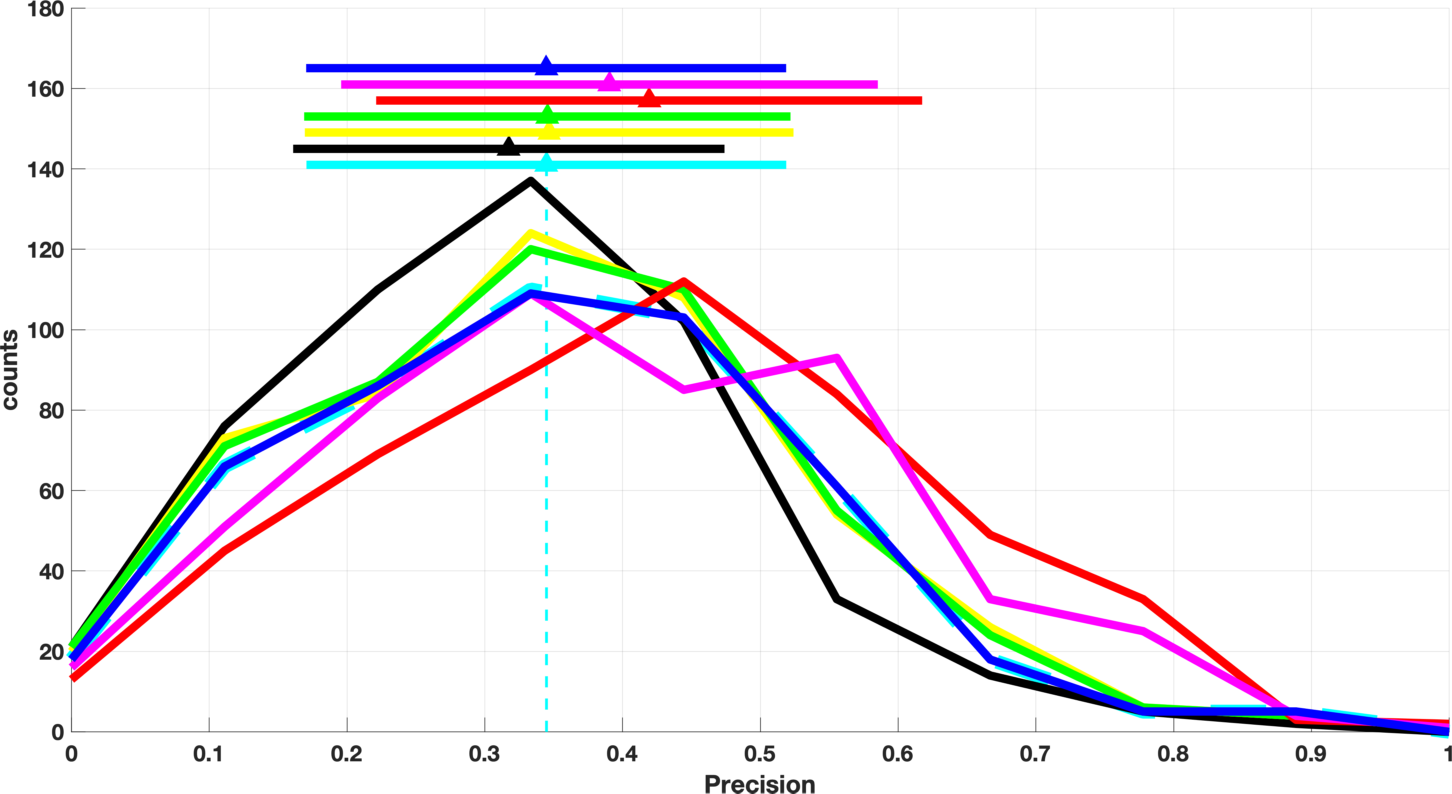} 
            \caption{Precision}
        \end{subfigure}
        \begin{subfigure}{.32\textwidth}
            \centering
            \includegraphics[width=\textwidth]{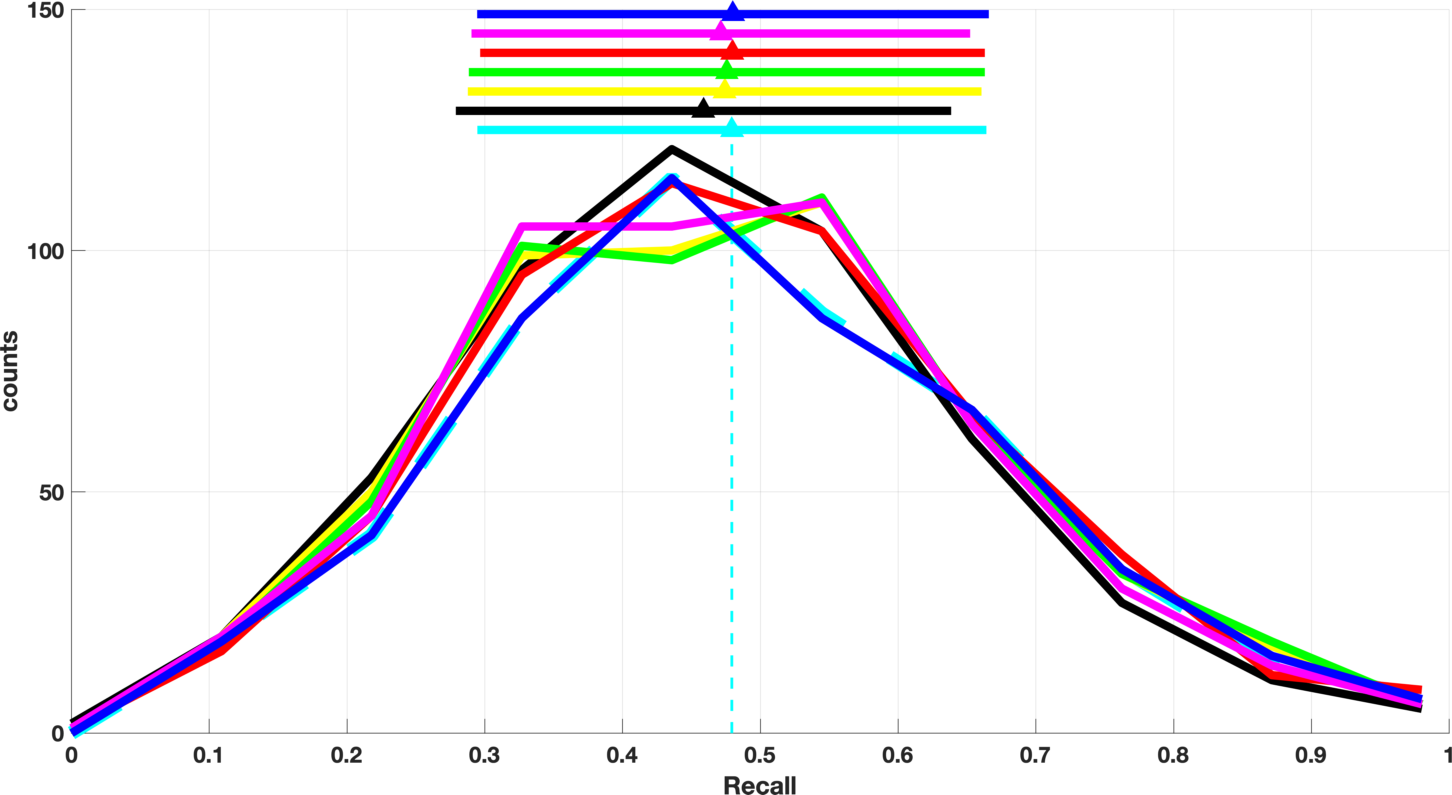}
            \caption{Recall}
        \end{subfigure}
        \caption{ \small{ \textbf{ Precision \& Recall model comparison:} (a) F-measure, (b) precision and (c) recall across 1000 image patches for Gaussian RF independent sensors baseline model and 4 network models with optimized parameters and $d_t=2$. Distribution $\mu$ and $\sigma$ denoted above.} Note colors same as in Figs.~\ref{KurVsEig}\&\ref{BlurVsNo}.}
        \label{PrecRec}
    \end{figure}

\begin{figure}[H]
        \centering
        \begin{subfigure}{.48\textwidth}
            \centering
            \includegraphics[width=\linewidth]{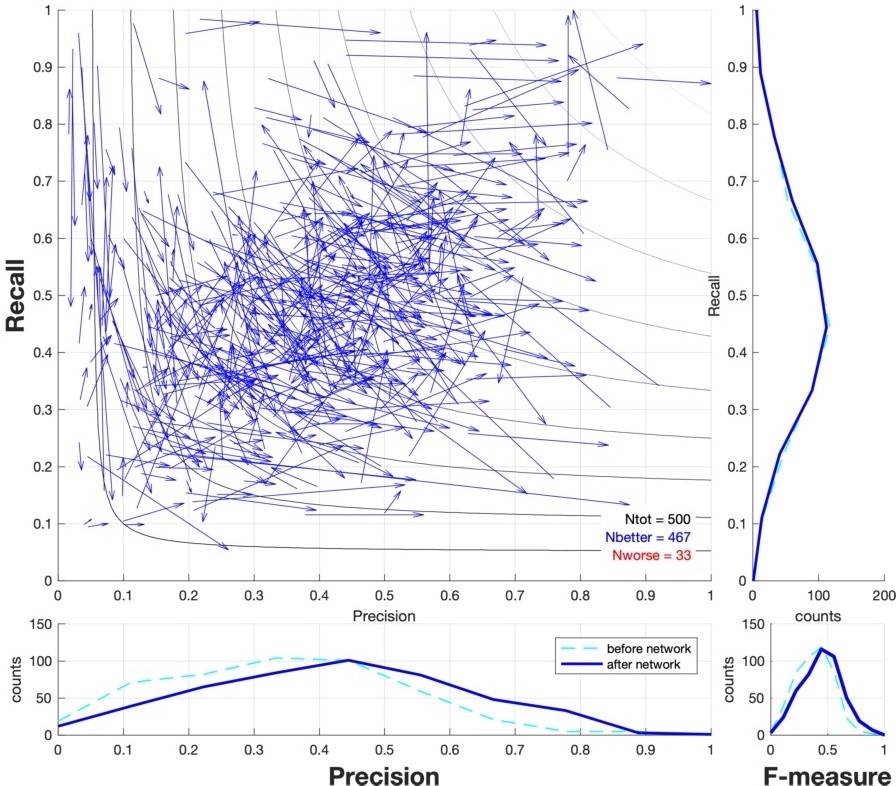}
            \caption{Improved segmentation: $ \Delta F > 0$}
        \end{subfigure}
        \begin{subfigure}{.48\textwidth}
            \centering
            \includegraphics[width=\linewidth]{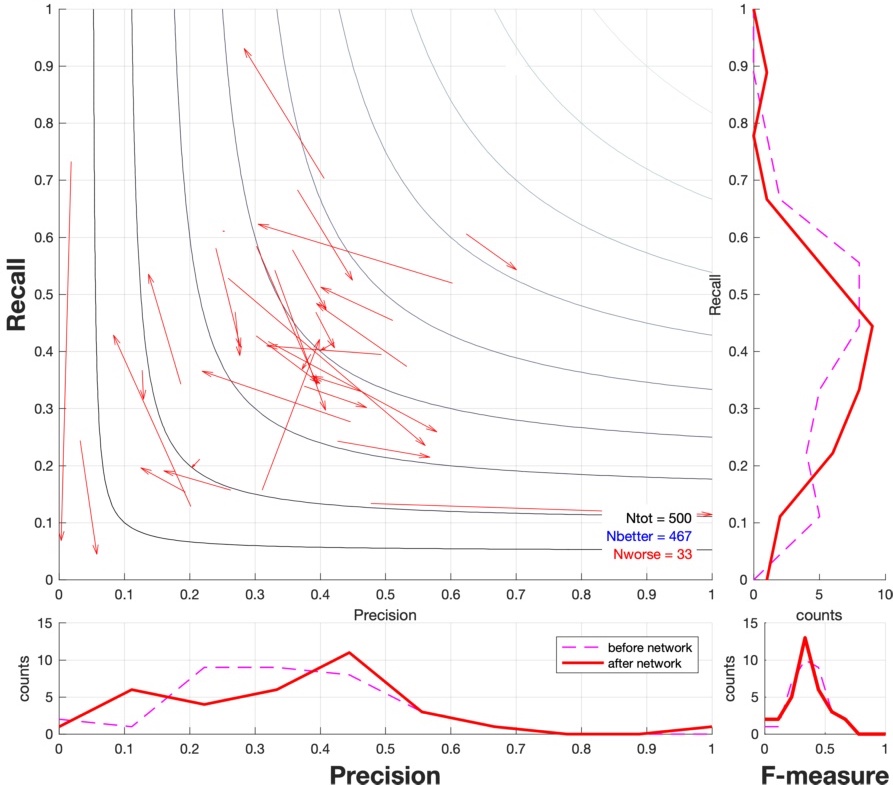}     
            \caption{Degraded segmentation: $\Delta F < 0$}
        \end{subfigure}
        \caption{ \small{ \textbf{ $\Delta$Precision and $\Delta$Recall with TM-2D model:} In \textit{ panel a}, arrows show change in P \& R for 467 image patches where network increased F-measure. Arrow tails indicate values before network relaxation and heads values after. Surrounding are distributions showing P,R,F before network (in cyan) and after (in blue). \textit{ Panel b} shows the same for 33 image patches where network decreased F-measure. Distributions before network in magenta. } }
        \label{PrecRecScatters}
    \end{figure}

    The F-measure combines the performance measures Precision and Recall, each with intuitive interpretations described in section \ref{BSDS}. To analyze the differences between our different models, we separately plot the precision and recall distributions in Fig.~\ref{PrecRec}. Note the position of curves for each network method relative to the independent sensors Gaussian RF baseline model (cyan dashed curve). Focusing first on the F-measure, in panel a, three of the network models (AA in yellow, GL in green, M in blue) did not show significant differences. The ISO model (black) degraded segmentation performance while the TM models (red \& magenta) improved relative to the Gaussian RF baseline. In panels b and c, the precision distribution of both TM models shifts significantly to higher values while the recall distribution shifts only slightly to lower values. Thus, the performance improvement of the TM model is mainly caused by increased precision, reflecting superior ability to suppress spurious boundaries, texture or "noise" in the probabilistic boundary maps. 
    To better understand the computation in the TM-2D model, we visualize changes to Precision and Recall together for individual image patches in Fig.~\ref{PrecRecScatters}. 
    The TM-2D network relaxation improved segmentation for $\sim93\%$ of all image patches, in blue, panel a. Clear positive shifts in the precision and F-measure distributions can be observed from the independent sensors Gaussian RF model (dashed cyan) to the phase output from the TM-2D network relaxation (solid blue). No clear trend emerges for the recall distribution with improved images. No clear trend exists for images where the network relaxation decreased performance. For some precision increased, and recall decreased. For others, vice versa.

    \subsection{Visual assessment of model performances} \label{Vis_ass}

\begin{figure}[H]
      \centering
      \centerline{\includegraphics[scale=0.4]{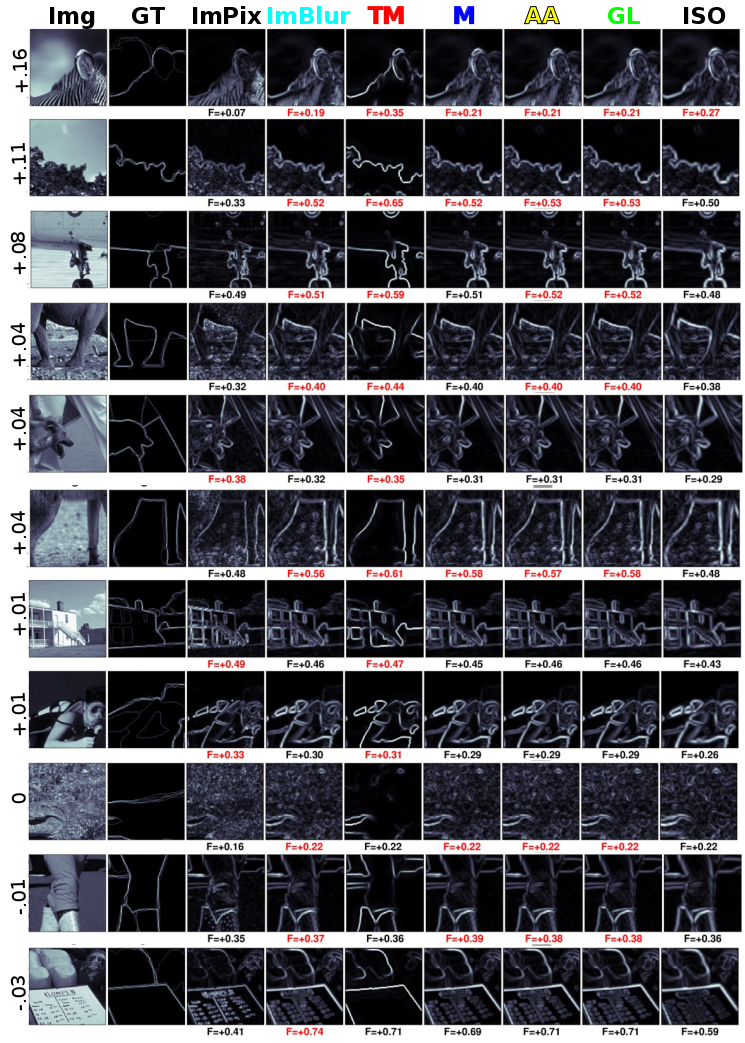}}
      \caption{ \small{ \textbf{Representative image patches:} Each row shows one example image patch ordered by change in F-measure between Gaussian RF independent sensors baseline and TM models (indicated on left).   Columns show image pixels, gT boundaries and pb maps obtained from raw pixels, Gaussian RF and 5 network models.  Mean F-measure value across all gT's noted below each pane is red if $\Delta F_G > 0$.} }
    \label{RepSamples}
    \end{figure}

    Finally, to provide some intuition what a $\Delta F$ value means for individual images, some examples are shown in Fig.~\ref{RepSamples}. Compared to the results from other methods, the TM model produces probabilistic boundaries (pb's) that are often thinner and cleaner. 
    
\begin{figure}[H]
        \centering
        \includegraphics[scale=0.25]{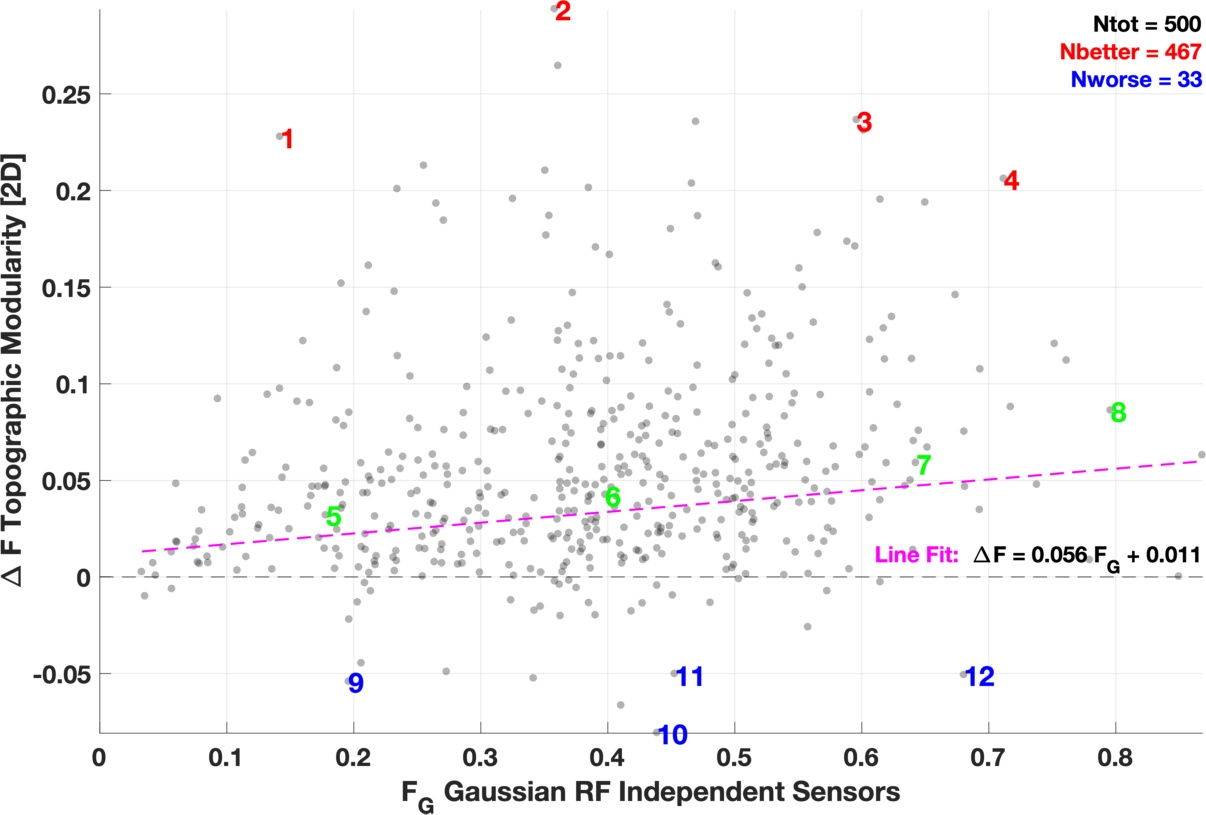}
        \includegraphics[scale=0.25]{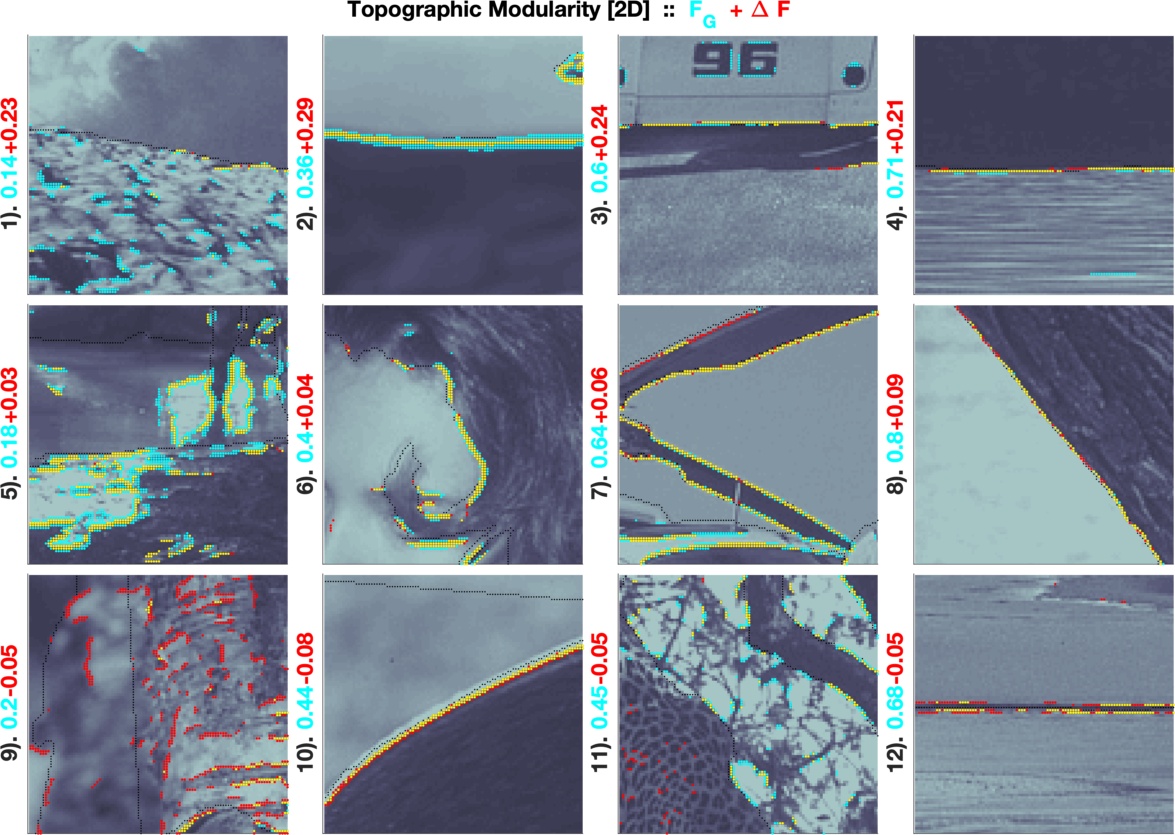}
        \caption{ \small{ \textbf{ Examples of TM 2D model performance:} \emph{Top} panel scatters F-measure in Gaussian RF independent sensors model vs. $\Delta$-F after 2D topographic modularity network phase relaxation. Out of 500 total image patches, 467 show positive improvement. Best fit line to scatter points in magenta. Colored numbers indicate randomly sampled image patches (shown in bottom panel) where $\Delta$-F performance is best (\#1-4), average (\#5-8) and worst (\#9-12). \emph{Bottom} panel shows image patches with best matching ground truth boundaries, in black. Yellow points indicate pixels found to be boundaries both by the Gaussian RF independent sensors model and the topographic modularity network model. Cyan number and points indicate F-measure under Gaussian RF model and boundaries found only by it. Red number and points indicate $\Delta$-F after TM 2D network phase diffusion and boundaries found only by TM-2D. Note that image patches are shown at $\nicefrac{1}{2}$ contrast to highlight boundaries found.} }
        \label{examples_better_worse}
    \end{figure}

Further, in Fig.~\ref{examples_better_worse}, we show samples of image patches with varying image segmentation performance relative to the Gaussian RF independent sensors model.

\section{Discussion}

    Here we have shown that phase relaxation in coupled oscillators receiving inputs from simple image sensors (with unoriented Gaussian receptive fields) can provide image segmentation performance above and beyond the baseline, the segmentation performance that can be achieved by just using local contrast measurements. 
    First, we have demonstrated that the type of graph clustering matters, the common spectral methods do not perform as well as relaxation in a Kuramoto model \cite{arenas2007}. Second, we have demonstrated that the graph derived from the image structure matters. Specifically, we introduced topographic modularity, a modularity matrix that can capture the distance dependence in the statistics of image features. We find that a Kuramoto model using the topographic modularity matrix as phase couplings was the only network model that significantly outperformed the baseline. 
    
\begin{figure}[H]    
          \centering
          \centerline{\includegraphics[scale=0.5]{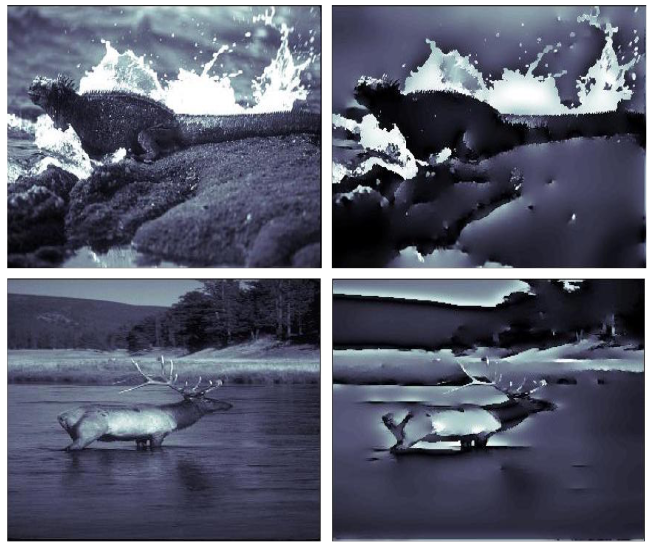}}
          \caption{ \small{ \textbf{ Two examples of cartoonization:} Original images on left and resulting phase of TM-2D network computation on right } }
        \label{Cartoonization}
    \end{figure}

    A critical ingredient in the successful model is negative phase coupling weights, which introduces phase desynchronization at segment boundaries. Interestingly, we saw the best segmentation results with Gaussian receptive fields sizes similar to those measured in retina \cite{croner1995}.
    In essence, the successful segmentation model provides a "cartoonization" \cite{yin2005} of images - smoothing texture and variation within segments while maintaining crisp segment boundaries. Examples of phase relaxation results on two sample images are shown in Fig~\ref{Cartoonization}. Note the halos at the base of the lizard tail and surrounding the elk, where low contrast segment boundaries have been accentuated.

    We quantify performance on the BSDS and show that anisotropic phase diffusion through the TM-2D improves F-measure significantly above baseline performance.  This improvement is obtained by increased Precision with only slightly decreased Recall. However, there are some caveats with benchmarking our retina model on the BSDS data. First, BSDS is designed for state-of-the art image segmentation methods that require combinations of sophisticated image filters, etc. However, context extraction in the retina can only use the simple image features of retinal cells, such as center surround features. Second, human image segmenters that provide the ground truth in the BSDS database can take advantage of the full image in color, while our model has only access to a $100 \times 100$ pixel image patch in greyscale. Third, humans segmenting images use consciously and unconsciously high-level semantic information to draw boundaries, while our algorithm just uses information from the image patch. 
    
    The model presented in this work is abstract and does not directly map to the biological features of retina. But some experimental evidence supports the plausibility of a computation in retina, as proposed by our model. Ganglion cell spike trains have been observed to be periodic in the Gamma frequency range \cite{neuenschwander1996} and the phase of this periodicity is transmitted with high precision through thalamus spikes to cortex \cite{koepsell2009}. The time to first spike in ganglion cells is quite precise \cite{gollisch2008} and provides a possible mechanism for phase initialization following global suppression during eye saccades \cite{roska2003}. 
    The phase coupling (without amplitude coupling) in our model could result simply from weak interactions between retinal cells, that slightly advance or delay spikes without adding or removing them. Both, phase synchronization and desynchronization through positive and negative weights in the model can be mapped onto excitation, inhibition and inhibition-of-inhibition circuits in retina. 
    The spatial null model's distance dependent term, $R_{ij}$ term in Eq.~\ref{eq_mod}, which requires global knowledge in the model could be implemented in retina via sampling through long distance inhibitory interactions from polyaxonal amacrine cells \cite{olveczky2003} or through eye movements implementing a temporal null model based on comparing feature similarity at a current stimulus location to feature similarity at a previous fixation. 
    However, one central feature in our model still lacks experimental support. The model requires a mechanism for fast adaptation of the phase couplings to a particular stimulus.
    
    Our modeling results suggest that, in principle, a coarse image segmentation or grouping/clustering of image features could be computed at the first stage of visual processing, in retina. While individual cell spike rates encode local stimulus contrast features through Gaussian-like receptive fields of ganglion cells, fine-time spike synchrony across the cell population encode extra-classical receptive field features, such as extended image segments. Fine-time correlations are multiplexed into ganglion cell spike-trains alongside with the rate-coded local stimulus features \cite{koepsell2010}. 

    \noindent  
    {\bf Acknowledgements:} CW has been supported by the Systems On Nanoscale Information fabriCs (SONIC) program, FTS has been partly supported by grant 1R01EB026955-01 from the National Institute of Health. Both researchers have been partially supported by NIH award 1R25MH109070-01.

\newpage

\bibliographystyle{abbrv}
\bibliography{Warner_PLOS_ImgSegRet.bib}

\appendix


\section{Optimal Gaussian RF size}
\label{Optimal_GaussRF}

There are multiple independent sensor models to which we could compare the network models. We constrain our sensors to have access to relatively simple image features similar to those which the retina would encode. For comparison, we compute image segmentation using two independent sensor null models. The first uses raw image pixels and the second passes image pixels through Gaussian filters that mimic retinal ganglion cell (RGC) receptive fields (RFs). Center-surround RGC RFs are modelled by a difference-of-Gaussian filter with an excitatory center and inhibitory surround. Gaussian filters fit to the centers and surrounds of primate midget and parasol ganglion cells were observed to be strongly center dominant \cite{croner1995}. Thus the receptive field of an RGC can reasonably be modelled by a single excitatory Gaussian center to first approximation and the optimal Gaussian RF size reasonably matches average RGC RF sizes measured in primate retina. 

In our simulations, the phase initialization of each individual oscillator as well as the connectivity strength between oscillators are both determined by the cell's activation - that is, how closely incoming stimulus matches the filter that is defined as a cell's receptive field.  We began with the simplest receptive field model, each cell responding to the greyscale pixel intensity value at its location.  Then, motivated by the biological fact that retinal receptive fields are spatially extended, we extended the receptive field model for each oscillating cell to be a localized Gaussian RF kernel.  To determine the best Gaussian RF size ($\sigma$), we numerically explored a range of spread values and kept the one that provided best average segmentation performance across ~500 image patches in the Berkeley Segmentation Dataset (BSDS) \cite{martin2001}.  Segmentation performance was determined by F-measure calculated  on the match between spatial gradients in phase maps output by network models and ground truth boundaries drawn by human subjects.  Interestingly, we determined that a Gaussian RF kernel with $\sigma=1$ pixel performed best empirically, improving the F-measure value by a modest but statistically significant 0.04 points over raw image pixels.

Motivated further by the excitatory and inhibitory center-surround nature of biological receptive fields in retina, we employ difference of gaussian (DoG) filters with parameters based on retinal physiology \cite{croner1995}.  The Croner paper provides parameters fit to DoG receptive fields for M and P cells in primate retina for eccentricites ranging from $0-40^{\circ}$ in its Table 1.  In contrast with LGN center-surround cells \cite{martinez2014}, retinal receptive fields have very weak surrounds ($\sim \nicefrac{1}{100}^{th}$) compared with the strength of the center portion.  From the many receptive field parameters fit to different cell types at different eccentricities in the primate retina, we distilled out 4 clusters that were different enough to test via simulations. In our simulations using DoG filters with P-avg and M-avg parameter values, we did not see image segmentation improvement over simple Gaussian filter with $\sigma=1$.

\begin{figure}[H]

    \centering
    \begin{subfigure}{.45\textwidth}
        \centering
        \includegraphics[scale=0.6]{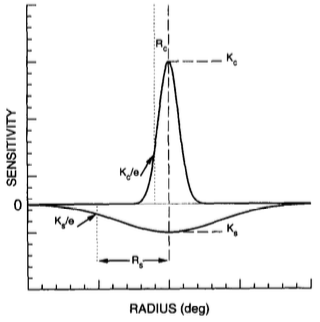}
    \end{subfigure}
    \hspace{2cm}
    \begin{tabular}{ c | c | c | c }
        & $R_c$ & $R_s$ & $\nicefrac{K_s}{K_c}$ \\ 
        \hline
         P-avg  & 1         & 8         & 0.01 \\ \hline
         P-$40^{\circ}$  & 3        & 13        & 0.06 \\  
         \hline
         M-avg  & 3        & 14.5     & 0.01 \\
         \hline
         M-$40^{\circ}$ & 5        & 12.5     & 0.025
    \end{tabular} 

    \caption{ \textbf{Primate center-surround RFs:} modeled as difference-of-Gaussians. Note: $R_c$ and $R_s$ in image pixels.  Values are given for magnocellular projecting (P) and parvocellular projecting (M) cells averaged across all eccentricities (avg) and at the visual periphery ($-40^{\circ}$) Image of measured retinal RF size from Croner 1995 \cite{croner1995} }.
    \label{cronerTable}
\end{figure}

\begin{figure}[!h]
  \centering
  \centerline{\includegraphics[scale=0.6]{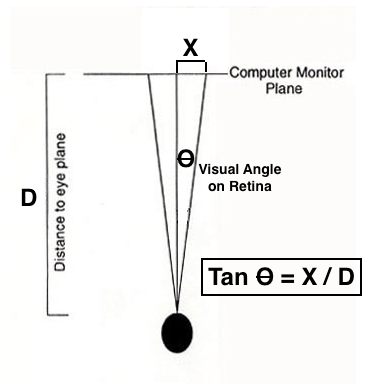}}
  \caption{ \textbf{Visual angle calculation schematic} }
\label{VisAng}
\end{figure}

Using a simple back-of-the-envelope visual angle calculation, illustrated in Fig.~\ref{VisAng}, and a few reasonable assumptions we approximate the size of retinal receptive field centers and surrounds in terms of image pixels for our models.  The calculation goes as follows:  Full images in the BSDS are 321 x 481 pixels and we assume that the displayed image size is 8.5'' x 11''.  Given these assumptions, an image pixel is approximately 0.02'' on a side.  Next, we assume that the projection screen is placed 24'' away from the eye.  Then, the angle that a single pixel subtends on the retina is approximately $0.05^{\circ}$.  Using this relation, we convert numbers provided in the Croner paper for retinal receptive field sizes into pixels and provide them in Fig.~\ref{cronerTable}.

\section{Motivating modularity} \label{MotivateModularity}





\subsection{Homogenization and Null Model as Expected Value of Weight}

Most generally, an entry in the \emph{modularity} matrix ($Q_{ij}$) is defined as the difference in weight between a pair of nodes in the actual network, characterized in the \emph{adjacency matrix} ($A_{ij}$), and the expected value of that weight ($\mathbf{E}[A_{ij}]$) in a ``homogenized network'', with connections between nodes made to reflect gross statistics of the network's connectivity.

\begin{equation}
	Q_{ij} = A_{ij} - \mathbf{E}[A_{ij}] \qquad \text{with} \qquad \mathbf{E}[A_{ij}] = \int A_{ij} p(A_{ij}) dA_{ij}
\end{equation}

The expected value of weights is parameterized in the \emph{null model} ($N_{ij}$) which is chosen to reflect the modeller's knowledge of network structure and connectivity.

\begin{equation}
	Q_{ij} = A_{ij} - N_{ij} \qquad \text{where} \qquad N_{ij} = \mathbf{E}[A_{ij}]
\end{equation}

The null model is constrained only by two considerations.  First, because the networks considered have undirected edges, both adjacency and null model matrices are symmetric, with $N_{ij} = N_{ji}$ and $A_{ij} = A_{ji}$. Second, it is axiomatically required that the total weight of edges in the null model are equal to the total weight of edges in the actual network because $Q=0$ when all the vertices are placed in the same partition.  This leads to a normalizing constraint on the null model matrix,

\begin{equation}
	 \mathbf{\Sigma} = \sum_{ij} A_{ij} = \sum_{ij} N_{ij}
	\label{constraint}
\end{equation}

\noindent where $\mathbf{\Sigma}$ is twice the total weight of edges in the network to account for double counting in the double sum over vertices (Note: $ \sum_{ij} := \sum{i} \sum{j} $). Beyond these basic requirements, we are free to choose from many possible null models, each one containing a different number of parameters, requiring a different number of computations and capturing the expectation of edge weights at different levels of homogeneity by calculating different statistics on the adjacency matrix.  


\subsection{I.I.D. or Homogeneous Random Graph}

The simplest null model, based on a Bernoulli or Erdos-Renyi random graph with weights allowed to take real values ( i.e. are not constrained to be binary), assigns a single uniform expectation weight to all edges in the network, $\bar{A}= \frac{\mathbf{\Sigma}}{n^2-n}$, which is the average edge weight in the actual network.  Note that $n$ is the number of nodes in the network and $ {n \choose 2} = \frac{n^2-n}{2}$ is the number of possible undirected edges that connect them with all-to-all connectivity, barring self-loops.  

\begin{equation}
	 \mathbf{E}[A_{ij} | \frac{\mathbf{\Sigma}}{n^2-n}] = \int A_{ij} \cdot p(A_{ij} | \frac{\mathbf{\Sigma}}{n^2-n}) dA_{ij} = \int A_{ij} \delta(A_{ij} - c \frac{\mathbf{\Sigma}}{n^2-n}) dA_{ij}
\end{equation}

\begin{equation}
	 N_{ij} = \mathbf{E}[A_{ij} | \frac{\mathbf{\Sigma}}{n^2-n}] =  c \frac{\mathbf{\Sigma}}{n^2-n}
\end{equation}

Solving for c by equation \ref{constraint}, we find

 \begin{equation}
 	c= \frac{n-1}{n}.
 \end{equation}

Combining the I.I.D. edge weight assumption with the constraint on total weight strength, we derive that the null model which assumes Bernoulli random graph connectivity patterns expects each weight in the network to take the following value.
 
 \begin{equation}
 	N_{ij} = \frac{ \mathbf{\Sigma} }{ n^2 }
 \end{equation}

This is a very simple representation of the network which requires only a single number - the average edge weight across the entire network ($\bar{A}$), however it is inadequate to capture the structure in all but the simplest networks.


\subsection{Independent-Vertex or Inhomogeneous Random Graph (N\&G Modularity)}

Relaxing the ``identical'' assumption of the I.I.D. graph null model, the ``Independent-Vertex'' model allows the expected value of each weight in the null model network to be different (inhomogeneous).  The expected value of a weight between two nodes is the product of the degree of each of those nodes.  This null model capture the expectation that two strongly connected nodes are more likely to be connected to one another and two nodes which are generally weakly connected are unlikely to be connected to one another.  Specifically, 

\begin{equation}
	 \mathbf{E}[A_{ij} | \frac{d_i}{n}, \frac{d_j}{n}] = \int A_{ij} p(A_{ij} | \frac{d_i}{n}, \frac{d_j}{n}) dA_{ij} = \int A_{ij} \delta(A_{ij} - c \frac{d_i}{n}\frac{d_j}{n}) dA_{ij}
\end{equation}

\begin{equation}
	 N_{ij} = \mathbf{E}[A_{ij} | \frac{d_i}{n}, \frac{d_j}{n}] = c \frac{d_i}{n}\frac{d_j}{n} \qquad \text{where} \qquad d_i = \sum_{i=1}^n{A_{ij}}
\end{equation}

\noindent where $n$ is the number of vertices and $d_i$ is the ``degree'' of node $i$ or strength of connectivity from node $i$ to all other nodes in the network, defined as the row (or equivalently column) sums of the adjacency matrix. Solving for c by equation \ref{constraint}, we find

 \begin{equation}
 	 c = \frac{n^2}{ \mathbf{\Sigma} }
 \end{equation}
 
\noindent making the full null model

\begin{equation}
	N_{ij} = \frac{d_i d_j}{ \mathbf{\Sigma} }. 
\end{equation}

 This requires $n$ numbers or statistics calculated from the network to characterize the null model, namely the degree of each node. This is the model used by Newman \cite{newman2006} and works well finding community structure in networks with no inherrent spatial layout or topography.     


\subsection{Line-Distance Dependent, Independent-Vertex Random Graph in 1D (Mod SKH Adj)}

In networks with 1D spatial relationships, where each vertex is more likely or more strongly connected to nearby vertices than to distant vertices, the independent-vertex null model which just considers vertex degrees fails to capture this spatial structure and the modularity's ability to find communities in such topographical networks suffers.  
The simplest spatial arrangement of nodes in a network is along a line in one dimension. Here, we can expand the vertex-independent null model to include a line-distance dependent ($b_{|i-j|}$) term which characterizes the expectation of a weight between nodes separated by a distance ($|i-j|$).

\begin{multline}
	 \mathbf{E}[A_{ij} | \frac{d_i}{n}, \frac{d_j}{n}, \frac{b_{|i-j|}}{n-|i-j|}] = \\ \int A_{ij} p(A_{ij} | \frac{d_i}{n}, \frac{d_j}{n}, \frac{b_{|i-j|}}{n-|i-j|}) dA_{ij} = \\ \int A_{ij} \delta(A_{ij} - c \frac{d_i}{n} \frac{d_j}{n} \frac{b_{|i-j|}}{n-|i-j|}) dA_{ij} 
\end{multline}

 \begin{equation}
 	N_{ij} = \mathbf{E}[A_{ij} | \frac{d_i}{n}, \frac{d_j}{n}, \frac{b_{|i-j|}}{n-|i-j|}] = c \frac{d_i}{n} \frac{d_j}{n} \frac{b_{|i-j|}}{n-|i-j|}
 \end{equation}
 
 \noindent where
 
 \begin{equation}
 	d_i = \sum_{i=1}^n{A_{ij}} \qquad \text{and} \qquad b_{|i-j|} = \sum_{k=1}^{n-|i-j|}{A_{k,k+|i-j|}}
 \end{equation}
 
 Solving for c by equation \ref{constraint} yeilds
 
 \begin{equation}
 	c = \frac{ n^2 \mathbf{\Sigma} }{ \sum_{ij} ( d_i d_j  \frac{b_{|i-j|}}{n-|i-j|} ) }
\end{equation}

\noindent and the full null model is

\begin{equation}
	N_{ij} = \frac{ d_i d_j \frac{b_{|i-j|}}{n-|i-j|} \mathbf{\Sigma} }{\sum_{ij} ( d_i d_j  \frac{b_{|i-j|}}{n-|i-j|} )}
\end{equation}

\noindent where $\frac{d_i}{n}$ is the average weight from node $i$ to other nodes in the network, and $\frac{b_{|i-j|}}{n-|i-j|}$ is the average weight between a pair of nodes separated by the distance $|i-j|$.  Since nodes are arranged along a line, their separation distance in 1 dimensional space directly translates into distance from the diagonal in the adjacency matrix.  Namely, the first off-diagonal contains weights between nodes separated by one distance unit, the second off diagonal by two units, and so on.  This method requires $2n$ values computed from $A$ to characterize the null model, the $n$ normalized row (or column) sums and the $n$ normalized diagonal sums.  Although it is not entirely correct for networks arranged on a 2D grid, it can be used and yeilds better performance than the Independent-Vertex null model.


\subsection{Grid-Distance Dependent, Independent-Vertex Random Graph in 2D (Mod SKH Euc)}

A more correct null model for networks constructed from images admits the arrangement of nodes in a 2D lattice.  The setup follows very closely the construction discussed above in the Line-Distance Dependent case with independent contributions from node degrees and from the connectivity-distance relationship across the entire network.  When nodes are arranged in a two dimensional grid, however, the relationship between distance in the network and location in the adjacency matrix is no longer simple to express mathematically, as in diagonal sums of $\mathbf{A}$ in the 1D case. Fig. \ref{2D_Bij} below shows entries in the adjacency matrix representing the collection of edges separating pairs of nodes by the distance indicated in each pane in an 11x11 image patch.  

\begin{figure}[!h]
  \centering
  \centerline{\includegraphics[scale=0.4]{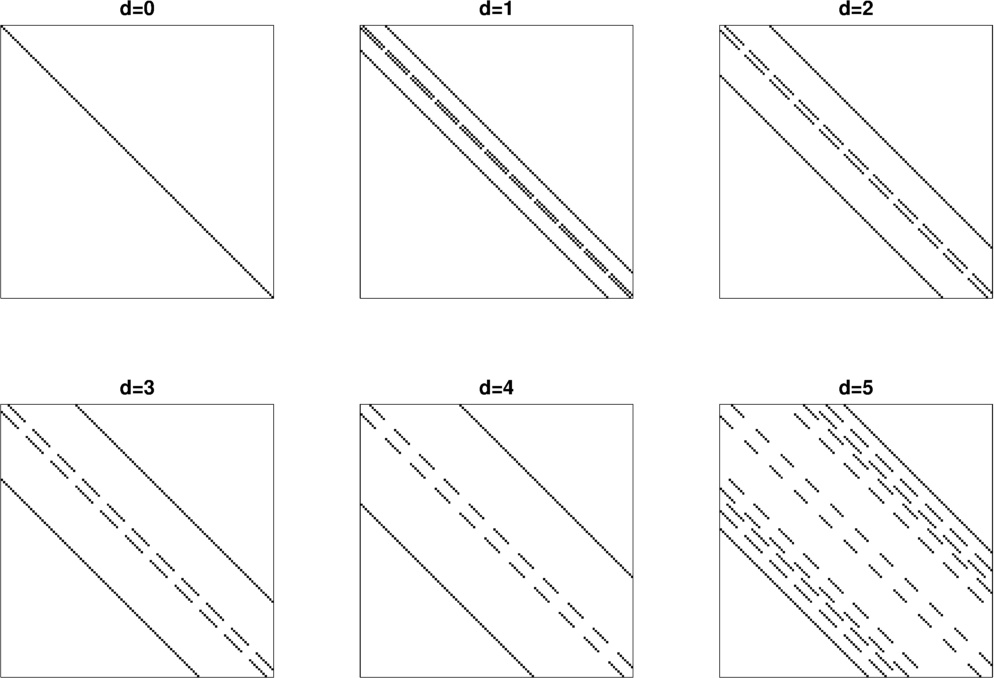}}
  \caption{ \textbf{Grid-Distance Dependence:} Distance mask in $\mathbf{A}$ matrix: Elements within the adjacency matrix that are separated by distance $d = |r_i-r_j|$ in an $11$x$11$ network arranged on a 2D lattice.}
  \label{2D_Bij}
\end{figure}

For all but $|r_i-r_j|=0$, distances in the image plane translate into patterns in the adjacency matrix that are more complex than just off-diagonals. Note that each pattern includes some of the $|r_i-r_j|^{th}$ off-diagonal, with additional entries resulting from the way which the $n$x$n$ image is rasterized to make to form the $n^2$x$n^2$ adjacency matrix. In our implementation, we do not attempt to express the $b_{|r_i-r_j|}$ term analytically, rather we algorithmically compute distances in the image plane and construct an adjacency matrix mask for each distance that we use to compute the distance-dependent average connectivity.  Aside from difference in implementation, the motivation behind this model is identical to the 1D case. Here specifically,

\begin{multline}
	 \mathbf{E}[A_{ij} | \frac{d_i}{n}, \frac{d_j}{n}, \frac{b_{|r_i-r_j|}}{\#b_{|r_i-r_j|}}] = \\ \int A_{ij} p(A_{ij} | \frac{d_i}{n}, \frac{d_j}{n}, \frac{b_{|r_i-r_j|}}{\#b_{|r_i-r_j|}}) dA_{ij} = \\ \int A_{ij} \delta(A_{ij} - c \frac{d_i}{n} \frac{d_j}{n} \frac{b_{|r_i-r_j|}}{\#b_{|r_i-r_j|}}) dA_{ij} 
\end{multline}

 \begin{equation}
 	N_{ij} = \mathbf{E}[A_{ij} | \frac{d_i}{n}, \frac{d_j}{n}, \frac{b_{|r_i-r_j|}}{\#b_{|r_i-r_j|}}] = c \frac{d_i}{n} \frac{d_j}{n} \frac{b_{|r_i-r_j|}}{\#b_{|r_i-r_j|}}
 \end{equation}
 
 where
 
 \begin{equation}
 	d_i = \sum_{i=1}^n{A_{ij}}
 \end{equation}
 
 and $b_{|r_i-r_j|}$ is implemented by masks illustrated in Fig. \ref{2D_Bij}. Here, the $\#b_{|r_i-r_j|}$ term refers to the number of non-zero entries in the mask for the given distance.  Since edges are undirected and $\mathbf{A}$ is symmetric, the distance mask could also be implemented using the upper or lower triangular version of the adjacency matrix.
 
 Solving for c by equation \ref{constraint} yields
 
 \begin{equation}
 	c = \frac{ n^2 \mathbf{\Sigma} }{ \sum_{ij} ( d_i d_j  \frac{b_{|r_i-r_j|}}{\#b_{|r_i-r_j|} } ) } 
\end{equation}

and the full null model with the normalization constant is

\begin{equation}
	 N_{ij} = \frac{ d_i d_j \frac{b_{|r_i-r_j|}}{\#b_{|r_i-r_j|} } \mathbf{\Sigma} }{\sum_{ij} ( d_i d_j  \frac{b_{|r_i-r_j|}}{\#b_{|r_i-r_j|} } )}.
\end{equation}


\subsection{Temporal Modularity Null Model} \label{temporal_null_model}

While topographic modularity models are powerful tools for image segmentation, it is difficult to interpret how they could be implemented in retinal circuitry.  The distance-dependent term $b_{|r_i-r_j|}$ requires that each edge in the network have access to global knowledge, namely the average edge weight across the entire network of all edges that span the same physical distance for the current input stimulus. However, the null model can constructed with only local information if each neuron pair samples and stores the average edge weight between them over an ensemble of past stimuli. Hebbian plasticity in the ganglion-amacrine cell anatomical connectivity network could nicely account for such a computation.

\begin{figure}[H]
  \centering
  \centerline{\includegraphics[scale=0.3]{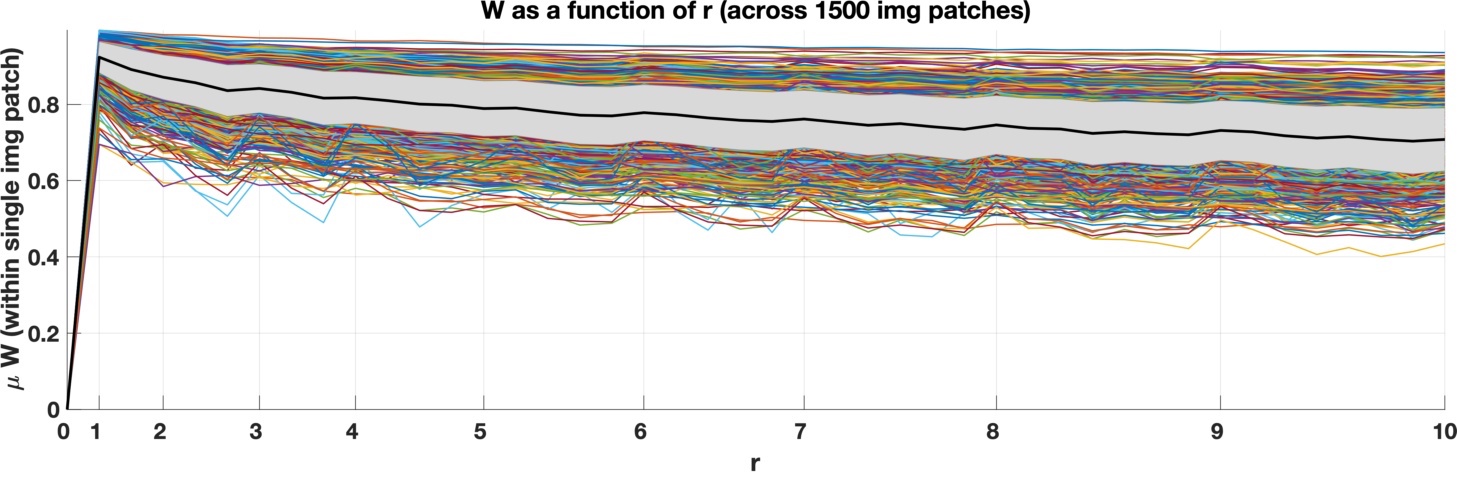}}
  \caption{\textbf{ Adjacency edge weight vs distance}: Average edge weight between node pairs in the adjacency matrix separated by distance r as a function of distance in image. Colored lines denote individual image patches and black line with grey error bars indicates $\mu$ and $\sigma$ across 1500 image patches that are 50x50pixels. }
  \label{Adj_vs_dist}
\end{figure}

Within a single scene or image, this spatial statistic can be converted to a local, temporal statistic via eye movements in a persistent scene if the timescale of plasticity is shorter than the scene duration \cite{zenke2017}. For longer Hebbian timescales, the argument holds across an ensemble of natural scenes in so far as the distance-dependent feature similarity in single images is captured by an average across the ensemble. Pixel values in images of natural scenes have been shown to be much more highly correlated for nearby pairs of pixels than for distant pairs \cite{atick1992}.Fig.~\ref{Adj_vs_dist} shows the average weight in the Adjacency matrix across all node pairs $i$ and $j$ separated by a distance $r = |r_i - r_j|$ as a function of $r$, within single image patches as colored lines and the mean and standard deviation across an ensemble in black and grey.

 A further advantage of a temporally sampled null model, beyond node degree and distance-dependence, is that \emph{all} parameters describing the relationship between cells (such as cell types and direction) are trivially captured the cell pair itself is used to compute the null model. Thus the null model effectively controls for all influences to network connectivity other than image content, which is marginalized out over many samples across time. The temporal null model has not been explored in this work and is left for future development.


\end{document}